\documentclass[12pt]{iopart}
\usepackage{iopams}
\usepackage{graphicx}
\usepackage{xcolor}
\usepackage{url}

\begin{document}

\title[Quantum Electrodynamics vacuum polarization solver]{Quantum Electrodynamics vacuum polarization solver}

\author{T. Grismayer$^1$, R. Torres$^1$, P. Carneiro$^1$, F. Cruz$^1$, R. A. Fonseca$^{1,2}$ and L. O. Silva$^1$}

\address{$^1$ GoLP/Instituto de Plasmas e Fus\~ao Nuclear, Instituto Superior T\'ecnico, Universidade de Lisboa, 1049-001 Lisbon, Portugal}
\address{$^2$ DCTI/ISCTE - Instituto Universit\'ario de Lisboa, 1649-026 Lisbon, Portugal}
\eads{\mailto{thomas.grismayer@tecnico.ulisboa.pt}, \mailto{luis.silva@tecnico.ulisboa.pt}}
\vspace{10pt}
\begin{indented}
\item[]May 2021
\end{indented}

\begin{abstract}
The self-consistent modeling of vacuum polarization due to virtual electron-positron fluctuations is of relevance for many near term experiments associated with high intensity radiation sources and represents a milestone in describing scenarios of extreme energy density. We present a generalized finite-difference time-domain solver that can incorporate the modifications to Maxwell's equations due to vacuum polarization. Our multidimensional solver reproduced in one-dimensional configurations the results for which an analytic treatment is possible, yielding vacuum harmonic generation and birefringence. The solver has also been tested for two-dimensional scenarios where finite laser beam spot sizes must be taken into account. We employ this solver to explore different types of laser configurations that can be relevant for future planned experiments aiming to detect quantum vacuum dynamics at ultra-high electromagnetic field intensities.
\end{abstract}

%
%
%
%
%

\section{Introduction}
\label{sec:intro}

The prospects offered by ultra-intense laser sources in the infra-red (IR) or X-ray central wavelengths~\cite{petawatt_review} have triggered a renewed interest in Quantum Electrodynamics (QED) and its impact on quantum processes at a macroscopic scale, namely how such phenomena can affect well studied interactions in the fields of plasma and laser dynamics. The most relevant QED processes in strong fields and high intensity laser interactions have been explored in several reviews~\cite{review_dipiazza, review_marklund, King_review}. Among these effects, the second order QED process of photon-photon scattering mediated by the vacuum fluctuation of virtual electron-positron pairs has been a topic of renewed interest motivated by several exotic consequences~\cite{review_marklund, HE_review, segev, matterless_slit} that originate directly from the original Heisenberg \& Euler Lagrangian~\cite{HE}. However, many of these effects, such as the virtual polarization of the vacuum, remain to be experimentally observed with the use of ultra-high intensity laser physics. With expected peak intensities up to $10^{23}-10^{24} $ Wcm$^{-2}$ to be delivered by large-scale facilities such as the Extreme Light Infrastructure (ELI)~\cite{ELI}, the VULCAN 10 PW project~\cite{Vulcan}, or the HERCULES laser upgrade~\cite{hercules}, the regime where these virtual fluctuations can be detected in the laboratory is close to being within reach. In particular, experiments are now being planned to study the quantum dynamics of the vacuum~\cite{Heinzel_experiment, HIBEF_Heinz} by combining ultra-intense optical lasers with X-ray lasers~\cite{HIBEF}. The increasing consensus regarding the importance of quantum dynamics in the collective effects of many extreme laser plasma interactions has motivated the development of novel numerical tools that couple the multiple scales associated with the problem. Numerical codes that simulate quantum radiation reaction~\cite{Marija_quantum, Green, Blackburn} and pair production effects~\cite{Elkina_rot, Vranic_merging, grismayer1, grismayer2, ridgers}, have already made important predictions in extreme energy density scenarios~\cite{ridgers, marija_RR, qed_pic_review, king_elkina}. We propose here an algorithm, different from the one suggested by Domenech \& Ruhl \cite{domenech}, that includes the effect of vacuum polarization in the spatio-temporal evolution of the electromagnetic field in multi-dimensions and allows for a broad set of initial conditions. Without dwelling deeply in the details of each approaches, both have advantages that deserve to be highlighted. Our approach is computationally less expensive and can be implemented in the PIC loop. On the other hand, the algorithm proposed by Domenech \& Ruhl \cite{domenech} is based on a gridless method which removes any interpolation issues and the precision of the numerical scheme is of fourth order. These vacuum quantum effects can be integrated via an effective nonlinear permeability and permittivity, and therefore to use a semi-classical approach. The effects of the quantum vacuum can be important and appreciable, not only in scenarios involving high intensity electromagnetic radiation, but also in extreme astrophysical environments surrounded by near critical Schwinger magnetic fields (e.g. neutron stars) where the propagation of electromagnetic waves is modified \cite{Heyl1, Heyl2}.

The electron-positron pair vacuum fluctuations were first taken into account by Heisenberg and Euler (HE) who calculated the first full Lagrangian to all orders \cite{HE}. In the low field $E \ll E_s$, low frequency  $\omega \ll \omega_c$ limit of the electromagnetic (EM) fields, the leading corrections of the standard Maxwell (M) Lagrangian density~\cite{HE} can be written as
\begin{equation}
\label{eq:lagrangian}
 \mathcal{L} = \mathcal{L}_\mathrm{M} + \mathcal{L}_\mathrm{HE} =  \frac{1}{4\pi}\mathcal{F}  + \xi( 4\mathcal{F}^2 + 7 \mathcal{G}^2 ), 
\end{equation}
where $\omega_c = m_ec^2/ \hbar$ is the Compton frequency, $E_{sch} = m_e^2 c^3/e \hbar$ the Schwinger critical field, and the EM invariants  $\mathcal{F} = \left(E^2-B^2\right)/2 $, $\mathcal{G} = \boldsymbol{E} \cdot \boldsymbol{B}$, $E$ and $B$ the electric field and magnetic field respectively. 
The nonlinearity coupling parameter is
\begin{equation}
\label{eq:xi}
 \xi = \frac{\alpha}{360 \pi^2 E_{sch}^2},
\end{equation}
with $\alpha = e^2/\hbar c$. The parameter weights the relative importance of the quantum corrections compared to the classical fields and vanishes in the  limit $\hbar \rightarrow 0$. From the Euler-Lagrange equations for the electromagnetic fields, we obtain a set of modified QED Maxwell's equations~\cite{segev}
\numparts
\begin{eqnarray}
\label{eq:faraday}
\boldsymbol{\nabla}  \cdot \boldsymbol{D} = 0 \\
\boldsymbol \nabla  \cdot \boldsymbol B = 0 \\
\boldsymbol \nabla  \times \boldsymbol H - \frac{1}{c}\frac{\partial \boldsymbol D}{\partial t} = 0   \\
\boldsymbol \nabla \times \boldsymbol E + \frac{1}{c}\frac{\partial \boldsymbol B}{\partial t} = 0 \textrm{ ,}
\end{eqnarray}
\endnumparts
with
\numparts
\begin{eqnarray}
\label{eq:D}
& \boldsymbol D = \boldsymbol E + 4\pi\boldsymbol P \\
\label{eq:H}
& \boldsymbol H = \boldsymbol B - 4\pi\boldsymbol M \textrm{ ,}
\end{eqnarray}
\endnumparts
and
\begin{equation}
\label{D_lagrange}
\boldsymbol{D} = 4\pi\frac{\partial \mathcal{L}}{\partial \boldsymbol{E}} \textrm{\ \ \ , \ \ } \boldsymbol{H} = - 4\pi\frac{\partial \mathcal{L}}{\partial \boldsymbol{B}}.
\end{equation}
The nonlinear vacuum polarization, $\boldsymbol P$, and magnetization, $\boldsymbol M$ read
\begin{equation}
\label{eq:P}
\boldsymbol{P} = \frac{\partial \mathcal{L}_\mathrm{HE}}{\partial \boldsymbol{E}} =   2\xi \left[ 2(E^2- B^2)\boldsymbol{E} + 7(\boldsymbol{E} \cdot \boldsymbol{B} )\boldsymbol{B} \right ]
\end{equation}
\begin{equation}
\label{eq:M}
\boldsymbol{M} = \frac{\partial \mathcal{L}_\mathrm{HE}}{\partial \boldsymbol{B}} =  -2\xi \left[(2(E^2-B^2)\boldsymbol{B} - 7(\boldsymbol{E} \cdot \boldsymbol{B})\boldsymbol{E} \right ].
\end{equation}
We would like to stress that we are using CGS units throughout the article. This semi-classical formulation treats the vacuum as a nonlinear medium, when the EM invariants $\mathcal{F}$ and $\mathcal{G}$ do not vanish. The algorithm aims to solve the nonlinear set of corrected Maxwell's equations leveraging the smallness of the vacuum non linearity. This algorithm described in this article is second order accurate in time and space. Due to its design, a key feature of the algorithm is that it can be seemingly incorporated into massively parallel fully relativistic electromagnetic particle-in-cell codes such as OSIRIS~\cite{OSIRIS}. This could also allow for studying self-consistently scenarios where charged particles are also present in the system, and to be explored in future publications.

This paper is organized as follows. In section~\ref{sec:2}, we describe the numerical algorithm, a generalization of the Yee algorithm, that solves Eqs.(\ref{eq:faraday}-\ref{eq:P}) in multi-dimensions. Section~\ref{sec:3} is devoted to the induced vacuum birefringence on an electromagnetic probe beam while crossing a region of high field. Several configurations of high field regions are considered such as a static field or realistic optical laser pumps. Harmonic generation constitutes the focus of section~\ref{sec:4} in one-dimensional and two-dimensional geometry. Finally in section \ref{sec:conclusions} we state the conclusions.

\section{Numerical Algorithm}
\label{sec:2}

\subsection{Description}

The standard second-order finite-difference time domain (FDTD) method to solve Maxwell's equations is the Yee Algorithm~\cite{yee}. The Yee scheme solves simultaneously for both electric and magnetic fields by solving Faraday's and Amp\`ere's law, respectively. The explicit linear dependence of Maxwell's equations on the fields allows the field solver to be centered both in space and time (leapfrog scheme), thus providing a robust, second order accurate scheme without the need to solve for simultaneous equations or perform matrix inversions \cite{taflove}. Moreover, the efficiency and simplicity of the Yee scheme allow an easy incorporation into numerically parallel codes. 

\begin{figure}
\centering
\includegraphics[width=.8\linewidth]{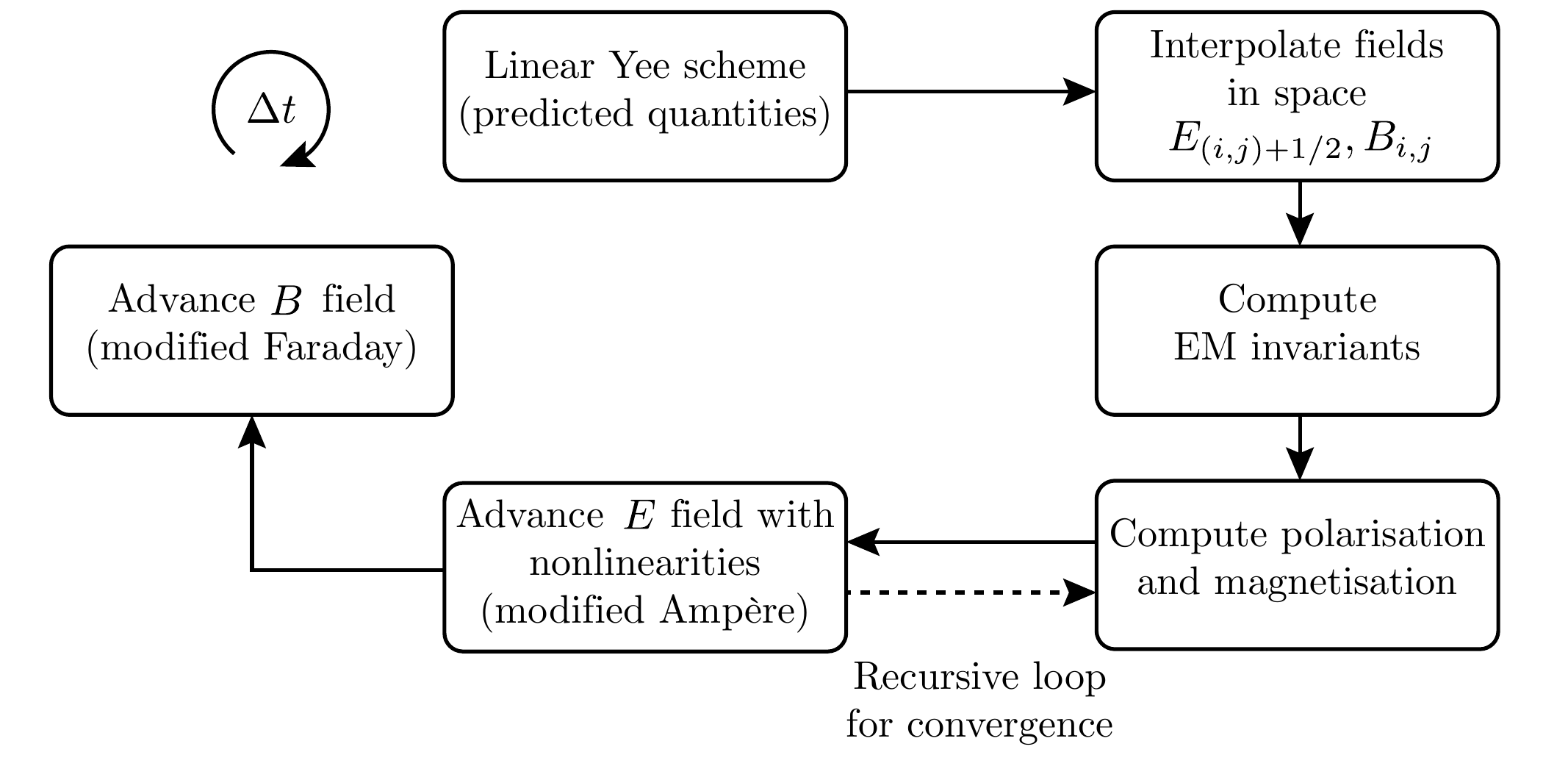}
\caption{Full loop of the modified Yee scheme}
\label{fig1}
\end{figure}

To solve the QED  Maxwell's equations, a modified Yee scheme was developed to address the two main difficulties which arise from the nonlinear terms. Firstly, all fields must be evaluated at all grid positions as opposed to spatially staggered fields. This permits to accurately evaluate quantities such as the EM invariants $\mathcal{F}$ and $\mathcal{G}$, since a given component of the nonlinear polarization and magnetization vectors now fully couples all other field components as can be understood from Eqs.(\ref{eq:P}-\ref{eq:M}). This is a significant obstacle regarding the essence of the Yee scheme since the algorithm may no longer be correctly spatially centered. Secondly, the temporal derivative of the nonlinear polarization term in Amp\`ere's equations prevents each electric field component to be advanced straightforwardly as it requires the knowledge of future quantities. This is easily understood through the discretization of the modified Amp\`ere's law in one dimension
\begin{eqnarray}
-\frac{B_{z \ i+1/2}^{n+1/2}-B_{z \ i-1/2}^{n+1/2}}{\Delta x} + 4\pi \frac{M_{z \ i+1/2}^{n+1/2}-M_{z \ i-1/2}^{n+1/2}}{\Delta x} = \nonumber \\ 
=\frac{1}{c} \frac{E_{y \ i}^{n+1}-E_{y \ i}^{n}}{\Delta t} + \frac{4\pi}{c} \frac{P_{y \ i}^{n+1}-P_{y \ i}^{n}}{\Delta t} \textrm{ ,}
\end{eqnarray} 
where the indices $n$ and $i$ denote the temporal and spatial positions, respectively. Usually, one would isolate the electric field term of temporal index $n+1$ to advance this field in time. However, to calculate this component one must know the polarization at time step $n+1$, which is a nonlinear function of all other fields at the new time step. The latter difficulty served as the main motivation to develop the modified Yee scheme proposed here. The numerical scheme is illustrated in \fref{fig1} for a time step $\Delta t$:
\begin{itemize}
\item[--] we begin by advancing the fields using the standard Yee scheme (i.e. without accounting for the polarization and magnetization of the vacuum). This setup allows us to obtain predicted quantities for the values of the fields at the new time. This approach is based on the standard technique of the predictor-corrector method, where the linear Maxwell's equations are solved as the zeroth order solution to the fields;
\item[--] the predicted field values are then interpolated at all spatial grid points using a spline interpolation method thus allowing to calculate quantities such as the EM invariants and respective polarization and magnetization of the vacuum, to lowest order;
\item[--] the polarization and magnetization are then used to advance the electric field via the modified Amp\`ere's law;
\item[--] the convergence loop re-injects this new electric field value back into the polarization and magnetization source terms Eqs.(\ref{eq:P}-\ref{eq:M}) to refine these quantities and re-calculate the electric field iteratively. This loop is reiterated until the electric field converges to a value within the desired accuracy;
\item[--] after convergence is achieved, Faraday's law is advanced one time step, identically to the linear Yee scheme, benefiting from the fact that the electric field values being used are self-consistent with the QED corrections.
\end{itemize}

It must be emphasized that this method is only valid as long as the effects of the polarization and magnetization of the medium are small compared to the non-perturbed propagation of the fields given as solutions to Maxwell's equations in classical vacuum. This condition is automatically satisfied for realistic values of electromagnetic fields available in current, or near future, technology. In this regime, the QED theory is valid since the Schwinger field, around which spontaneous pair creation (Schwinger effect) becomes non-negligible, corresponds to an electric field of $E_s \sim  10^{18} $V/m, whereas ambitious laser facilities aim to push available intensities to the $10^{23} - 10^{24}$ W/cm$^{2}$ ($E \sim  10^{-3} E_{sch}$) range. The order of the $\xi$ parameter in Eqs.(\ref{eq:P}-\ref{eq:M}) which is on the order $10^{-6}E_{sch}^{-2}$ clearly helps to ensure the validity of the method. Therefore, this scheme highly benefits from the fact that the nonlinear QED corrections of the vacuum are perturbative in nature. The convergence loop can be seen as a Born-like series since for every re-insertion of the fields back into the nonlinear source term, there is a gain in accuracy of one order in the expansion parameter to the result. The algorithm proposed here solves Amp\`ere's law by treating the nonlinear corrections as a source term, in an iterative manner,
\begin{equation}
\label{eq:ampere_source}
\boldsymbol \nabla \times \boldsymbol B - \partial_t \boldsymbol E = \boldsymbol S_\mathrm{NL}[E,B]\textrm{,}
\end{equation}  
where $\boldsymbol{S}_\mathrm{NL} = \boldsymbol{\nabla} \times \boldsymbol{{M}} + \partial_t \boldsymbol{{P}}$.
From this discussion and equation~\eref{eq:ampere_source}, we can conclude that this generalization of the Yee scheme can be extended beyond the framework of QED corrections to the vacuum as it is valid to solve Maxwell's equations in any nonlinear medium provided that the polarization and magnetization are given and that their order is such that they can be treated as a perturbation. This possible generalization enhances the range of applicability of our algorithm. Furthermore, the inclusion of a current in the algorithm ($J \neq 0$ in Amp\`ere's law) can be done, both within a PIC framework or for a macroscopic field dependent current by including the current term in the initial standard Yee scheme loop where the predictor quantities are computed. This is another key feature regarding the ability to couple our proposed generalized Yee solver to the PIC framework, of paramount importance to model scenarios where charged particles (even in small numbers) are present. 

\subsection{Interpolation of the fields}

\begin{figure}
\centering
\includegraphics[width=.8\linewidth]{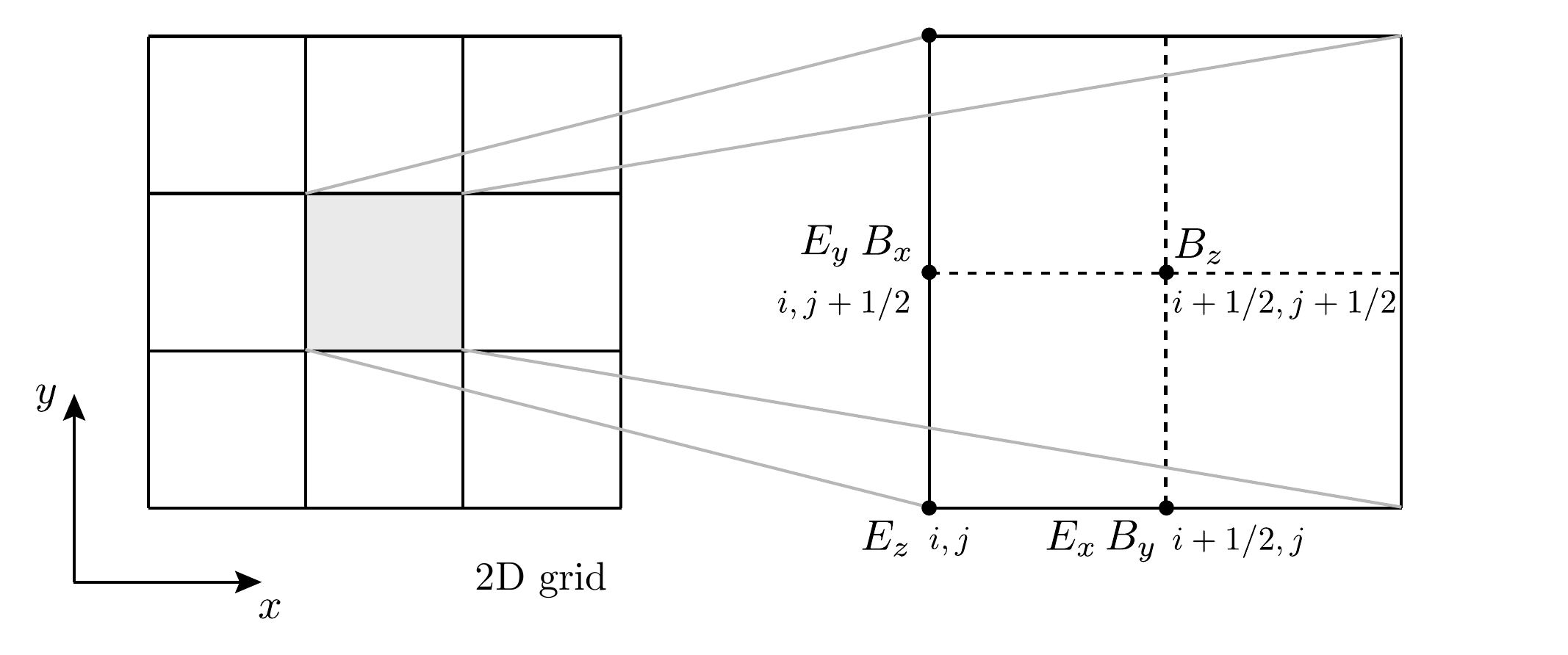}
\caption{Position of the electric and magnetic field vector components within a 2D cell of the Yee lattice.}
\label{fig2}
\end{figure}

The algorithm requires that all fields are calculated at the same spatial positions. When considering the spatial interpolation of the self-consistent fields given by the Yee Algorithm we found a clear asymmetry between interpolating the electric field at the magnetic field position or vice-versa in terms of the precision of the EM invariant $E^2-B^2$ for both cases. Since a plane wave is an exact solution of the QED Maxwell's equations, the invariants calculated in the simulation should be identically zero~\cite{Mckenna}. \Fref{fig2} shows the distribution of the EM fields within a two-dimensional Yee grid cell. We found that all the standard interpolation schemes yield invariants with much greater precision at the lower left corner of the cell compared to the other positions. This difference in precision was found to be of two orders of magnitude when tested for a plane wave in 1D, which can affect the stability and precision of the code. The reason for this artefact is due to the way that the fields are initialized within the simulation domain. In particular, the fact that the electric and magnetic fields must be initialized with a shift both in space and time, creates an asymmetry between interpolating a field to the corresponding position of the other field, even if this interpolation is done in a centered manner.  The solution we have adopted to address this problem is to calculate all the fields at the cell corner where the invariants are known to be of higher precision. For instance, the $B_z$ component at the left corner of the cell becomes
\begin{equation*}
B_{z \ i,j}=\mathcal{I}(B_{z \ i+\frac{1}{2},j+\frac{1}{2}},B_{z \ i-\frac{1}{2},j+\frac{1}{2}},B_{z \ i+\frac{1}{2},j-\frac{1}{2}},B_{z \ i-\frac{1}{2},j-\frac{1}{2}}),
\end{equation*}
where $\mathcal{I}$ is an interpolation function. Once all fields are calculated at the $(i,j)$ positions, we can compute the invariants at these positions and then re-interpolate these invariants directly to the other grid cell points in a similar fashion. The correct calculation of the EM invariants is necessary in order to evaluate the nonlinear polarization and magnetization of the vacuum via equations~\eref{eq:P}-\eref{eq:M}.

\subsection{Numerical Stability}
The method adopted to study the numerical stability of the QED polarization solver follows the standard mode analysis \cite{taflove}. With the linear Yee scheme, the one-dimensional numerical dispersion relation for a plane wave propagating on a grid with spatial and temporal resolution $\Delta x$ and $\Delta t$ respectively is \cite{taflove}
\begin{equation}
\label{linear_omega}
\omega = \frac{1}{\Delta t} \arccos \left( 1+ \left( \frac{c\Delta t}{\Delta x} \right)^2 (\cos(k \Delta x) -1) \right) .
\end{equation}
A notable case is when $\Delta t = \Delta x/c$ for which equation~\eref{linear_omega} reduces to the EM dispersion relation for a plane wave in vacuum, $\omega = kc$. To study the stability of the new set of QED-corrected Maxwell's equations using this method, a self-consistent numerical dispersion relation was derived. Due to the nonlinearity of the equations, the new numerical dispersion relation can be written as
 \begin{equation}
 \label{nl_omega}
 \left(\frac{c \Delta t}{\Delta x} \right)^2 \sin\left(\frac{k \Delta x }{2}\right)^2 - \sin\left(\frac{\omega \Delta t }{2}\right)^2  = \xi E_0^2F_\mathrm{NL}(\omega , k, \Delta x , \Delta t),
 \end{equation}
where $E_0$ is the amplitude of the wave and $F_\mathrm{NL}$ is a nonlinear function of $\omega$, $k$, and the spatial and temporal steps. In the classical limit $\xi \rightarrow 0  $ the RHS goes to zero and the dispersion relation reduces to equation~\eref{linear_omega}. A numerical plane wave propagating via our QED solver will therefore obey equation~\eref{nl_omega}.

\begin{figure}
\centering
\includegraphics[width=.8\linewidth]{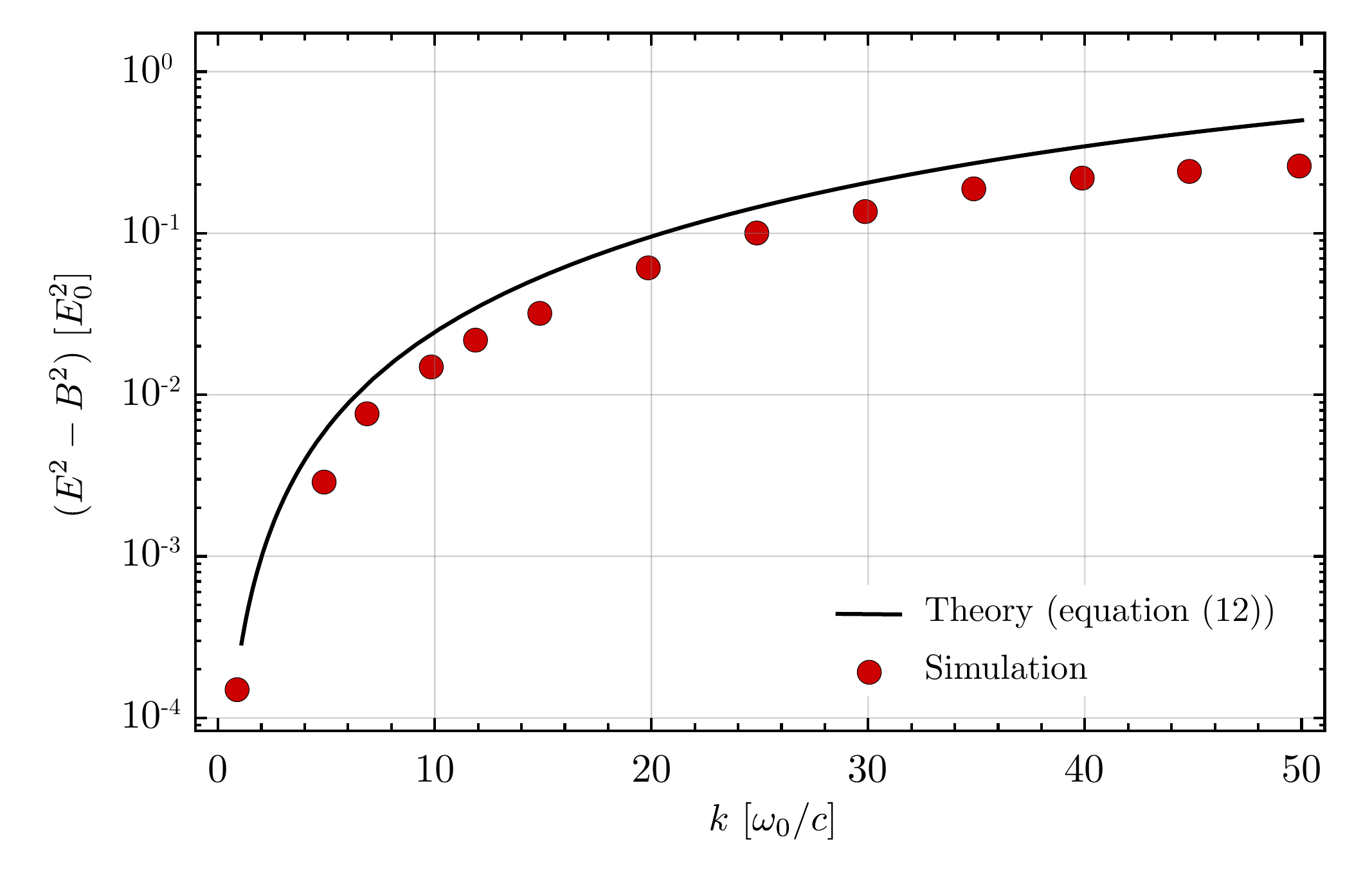}
\caption{Amplitude of EM invariant $E^2-B^2$ as a function of the seeded $k$ mode for a resolution $\Delta x =  \pi / 100$, $\Delta t = 0.98\Delta x $ and $\xi E_0^2 =  10^{-4}$. Simulation results in blue are compared to equation~\eref{invariants_fourier} in red.}
\label{fig:1invariants_1}
\end{figure}

For a numerical plane wave the EM invariant $\boldsymbol E \cdot \boldsymbol B$ is identically zero, whereas the invariant $E^2-B^2$ will not vanish identically due to finite spatial resolution and the fact that the fields must be interpolated in space to evaluate the invariants, as already discussed above. Therefore, the amplitude of this EM invariant depends on the interpolation method and grid resolution. We calculate this dependence by evaluating $E^2-B^2$ at a given grid point, taking into account that a correct centering in space implies that one of the fields must be interpolated to the position of the other (in this case the $B$ field, using linear interpolation). This yields
\begin{equation}
E^2-B^2 =   E_0^2 \left[ 1 - \frac{\sin^2(k\Delta x/2)}{\sin^2(\omega \Delta t/2)} \cos^2 \left( \frac{k\Delta x}{2} \right) \right].
\label{invariants_fourier}
\end{equation}
This expression was compared to the results extracted from one-dimensional simulations. \Fref{fig:1invariants_1} shows a comparison between equation~\eref{invariants_fourier} and simulations with several seeded $k$ modes and with $\xi E_0^2 = 10^{-4}$ , $c\Delta t = 0.98\Delta x $ and $k\Delta x =  \pi / 100$.
The simulation points agree with the trend presented by the theoretical curve. This result shows that  equation~\eref{invariants_fourier} provides an upper bound to the interpolation error when seeding a particular $k$ mode. In particular, the results show that for higher wave numbers, up to the resolution limit, the order of magnitude of the invariant amplitude increases, tending towards unity. One shall therefore limit the simulations to low $k$ modes to ensure the smallness of the invariants. 

\begin{figure}
\centering
\includegraphics[width=.8\linewidth]{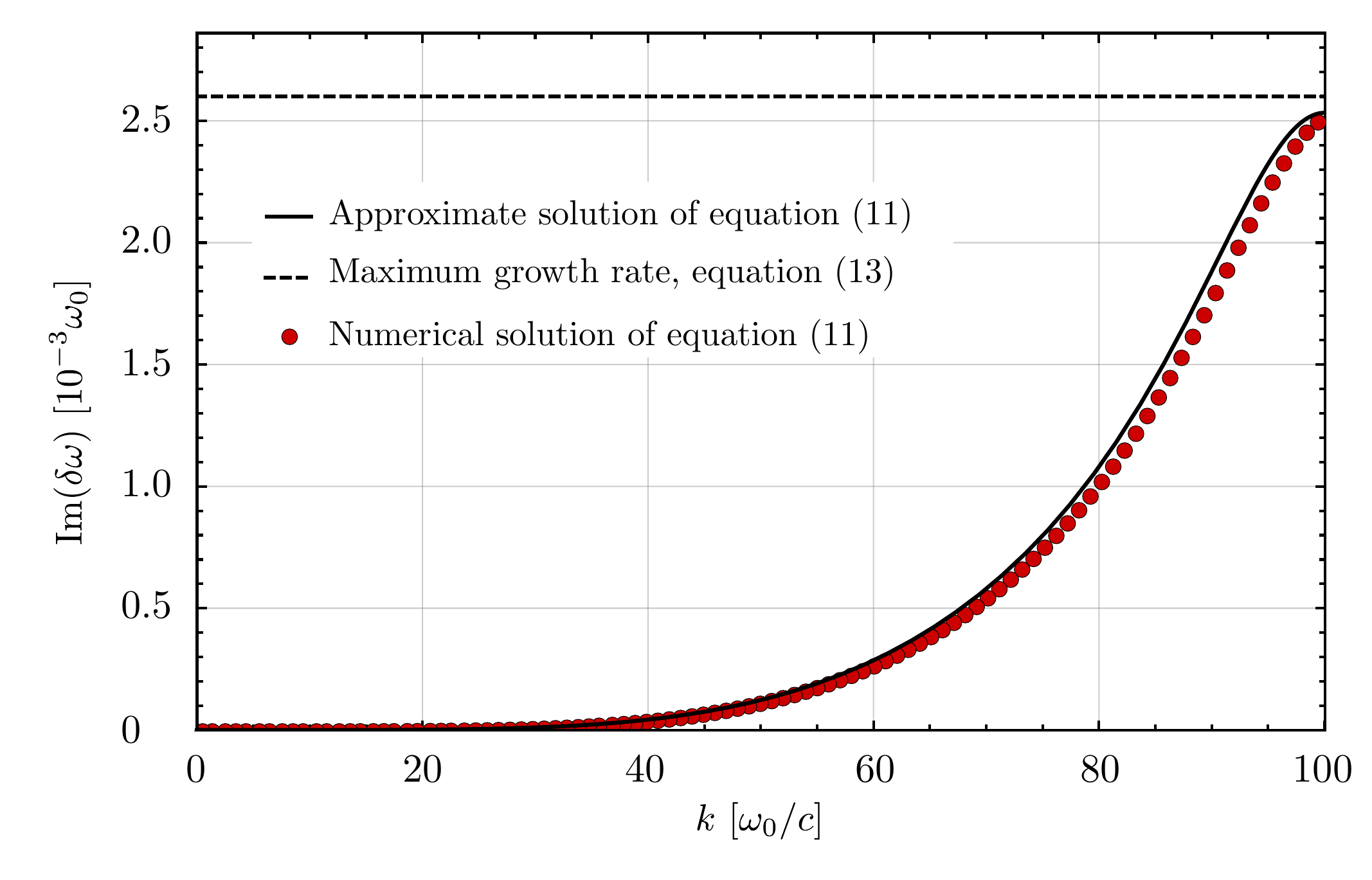}
\caption{Imaginary part of solution of nonlinear dispersion relation, equation~\eref{nl_omega}, as a function of $k$ mode, calculated using three different methods. Simulation parameters used were $ \Delta x = 0.0314 $, $\Delta t = 0.98\Delta x $ and $\xi E_0^2 =  10^{-4}$.}
\label{fig:3growhtrate0}
\end{figure}

\begin{figure}
\centering
\includegraphics[width=.8\linewidth]{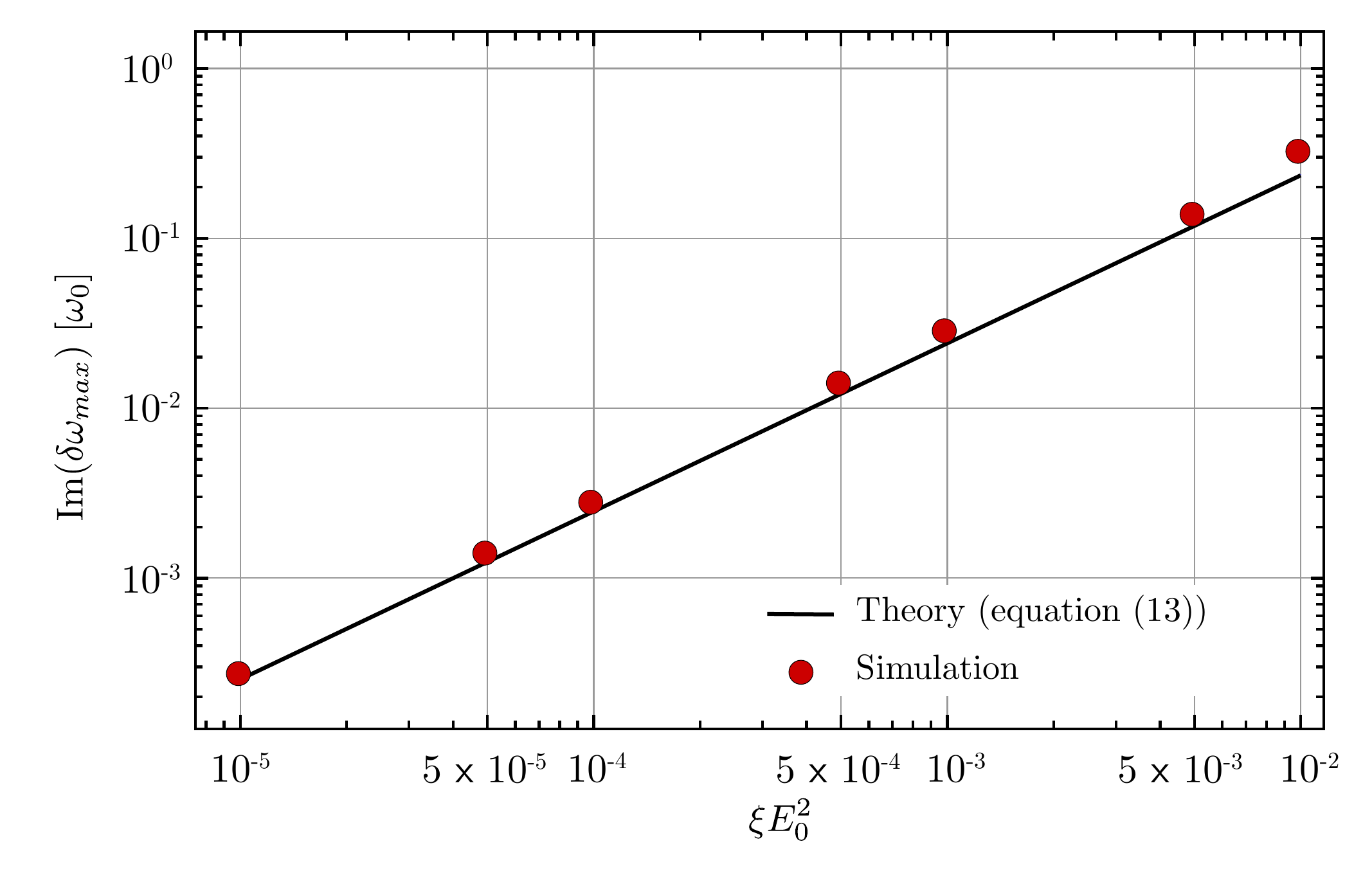}
\caption{Comparison between maximum growth rate extracted from simulation with theoretical prediction calculated from the nonlinear dispersion relation. The figure shows how this growth rate varies as a function of $\xi E_0^2$, which measures the importance of the nonlinear quantum vacuum corrections. The simulation parameters used were $ \Delta x = 0.0314, \Delta t = 0.98\Delta x$ and a seeded $k_{\mathrm{seed}} = 1$.}
\label{fig:7simulation_growthrate}
\end{figure}

The stability of the QED Yee solver, i.e., the nonlinear dispersion relation, equation~\eref{nl_omega} was solved using three methods: a numerical solution, an analytical solution through the linearization of the system via the ansatz $\omega = \omega_0+ \delta \omega$ with $\delta \omega \ll \omega_0 $, and finally by estimating the growth rate of the maximum mode allowed by the grid resolution i.e. $k\Delta x  = \pi $. The results are shown in \fref{fig:3growhtrate0} for simulations performed with a grid resolution of $ \Delta x = 0.0314 $, $\Delta t = 0.98\Delta x $ and $\xi E_0^2 =  10^{-4}$.
One can verify in \fref{fig:3growhtrate0} that the analytic solution is in excellent agreement with the numerical integration. The maximum growth rate is given by
\begin{equation}
\mathrm{Im}(\delta \omega_{max}) \simeq	 \frac{2\xi E_0^2 \sqrt{8 \varepsilon }}{\Delta t},
\label{Eq:7.approx_omega}
\end{equation}
where $\varepsilon = 1 - c\Delta t / \Delta x$. The maximum growth  rate predicted theoretically by equation~\eref{Eq:7.approx_omega} serves, therefore, as an accurate rule-of-thumb criterion to understand how unstable a given simulation setup may be. Finally we took the solution of the perturbative expansion and studied the limit for small $k$ values, which yields
\begin{equation}
\lim_{ k \Delta x \rightarrow 0} \mathrm{Im}(\delta \omega) = \frac{1}{4}\frac{\xi E_0^2 (k \Delta x)^5}{\Delta x} .
\label{small_k}
\end{equation}
Equation~\eref{small_k} suggests that the smallest $k$ modes will be the least unstable as not only does the growth rate scales with the small quantity $\xi E_0^2$, but also due to the power law applied to the small value of $k \Delta x$. This is an important result since, in principle, the low $k$ modes, for a given grid resolution, are those that will be seeded for a simulation setup.  
These theoretical predictions were compared with one-dimensional simulations by extracting the growth rate of a given $k$ mode in the simulation domain, the results are shown in \fref{fig:7simulation_growthrate}. The growth rate was extracted from the Fourier spectrum of the simulations for different values of $\xi E_0^2$. \Fref{fig:7simulation_growthrate} shows a good agreement between the maximum growth rates extracted and equation~\eref{Eq:7.approx_omega}.
Our theoretical analysis shows that the growth rate of the most unstable mode scales linearly with $\xi E_0^2$. Furthermore, we performed simulations under the same conditions, varying only the seeded $k$ mode and verified that this does not affect the growth rate of the most unstable high $k$ modes. Instead, it is the amplitude of the seeded mode that affects the growth rate of the higher $k$ modes by nonlinear coupling. It is possible to derive a criterion for the time at which the seeded field starts to be strongly deteriorated by the growing numerical noise, by assuming this blow-up occurs once the amplitude of the fastest growing $k$ modes $\delta \tilde{E}$ (initially this amplitude is at the numerical noise level, and can be measured from the initial spectrum of the fields in the simulation), become of the order of the initial seed amplitude. This criterion yields,
\begin{equation}
t_{\mathrm{blow}} \sim \frac{\Delta t}{\sqrt{\varepsilon}} \frac{1}{\xi E_0^2} \log \left( \frac{E_0}{\delta \tilde{E}} \right). 
\end{equation} 
For realistic values of $\xi E_0^2 $, this time is far greater than any simulation setup one may wish to perform.

\section{Vacuum Birefringence}
\label{sec:3}

A thorough benchmark of the functionality and robustness of the algorithm may only be gained by comparing simulation results with analytical results in 1D simplified cases. One-dimensional scenarios provide excellent opportunities to test the code against analytical predictions. The two cases we present here are the vacuum birefringence in the presence of a strong static field and counter propagating plane waves. Whilst the first case is well studied in the literature \cite{Baier, Brezin}, the second case requires a finer analytical work, yielding nevertheless the well known result of generation of higher harmonics due to the nonlinear interaction as shown in \cite{King1, fedotov, Dipiazza_HHG} for different setups and physical regimes that will be addressed in the next section. The true physical value of the parameter in normalized units of the simulation is $\xi \sim 10^{-17}$ for an optical frequency. Simulations were performed with artificially increased values of the $\xi$ parameter to better illustrate the method proposed here. This does not alter the physical relevance of the results. Rather, this is simply a re-scaling of a constant to highlight the effects more clearly. Finally, in all the results presented in this section, the units were normalized to the characteristic laser frequency, $\omega_0$ and wave number used in the simulations, $k_0$. The normalizations are thus $ t \rightarrow \omega_0 t$ and $x \rightarrow k_0x$. These normalizations of space and time define the normalizations used for the fields, i.e: $E \rightarrow eE/mc\omega_0 $ and $B \rightarrow eB/mc\omega_0$.

\subsection{Vacuum birefringence with a static field}

The birefringence of the vacuum is a thoroughly studied setup of great experimental interest to explore the properties of the quantum vacuum~\cite{plvas_birefringence, BMV}.   
A one-dimensional wave packet traveling in the presence of a strong static field will experience a modified refractive index of the vacuum due to the HE corrections. To obtain an approximate analytical expression, one assumes that the strong background field remains unperturbed by the nonlinearities. This motivates the following ansatz for the solution of modified Maxwell's equations
\begin{equation}
\label{eq:static}
\boldsymbol{E} = \boldsymbol{E}_p(x,t) + \boldsymbol{E_s},
\end{equation}
where $\boldsymbol{E}_p$ and $\boldsymbol{E}_s$ represent the electromagnetic pulse and the static fields, respectively and $E_p \ll E_s$. Inserting equation~\eref{eq:static} into the QED Maxwell's equations and keeping only the dominant terms in the polarization and magnetization, one obtains the following refractive indices:
\begin{eqnarray}
\label{npar}
n_{\parallel} &=&  \left ( \frac{1+6\xi E_s ^2}{1+2\xi E_s ^2} \right )^{1/2}, \\
\label{nperp}
n_{\perp} &=&  \left ( \frac{1+2\xi E_s ^2}{1-5\xi E_s ^2} \right )^{1/2},
\end{eqnarray}
where the parallel and perpendicular directions refer to the direction of the probe polarization when compared to the static field. 
In the case of a constant externally imposed magnetic field, the expressions of the indices are swapped \cite{Brezin} (the value of the perpendicular index takes the value of the parallel index and vice-versa). Notice that the product $\xi E_s^2$ appears as a relevant quantity. This is a recurring property of several setups. It must be ensured that this product is a small quantity, both for the validity of the theoretical framework but also from the  point of view of the algorithm. This quantity controls whether the corrections to the unperturbed fields are small or not, a crucial feature for the stability of the algorithm as already discussed. It is worth mentioning that the effective vacuum refractive indices only depend here on the externally imposed field to the order considered. Bialynicka-Birula \cite{BIALYNICKABIRULA} showed that if one considers the HE Lagrangian to all orders, a non collimated wave packet will experience, in addition to the external field effects, self-interaction. The dependencies of the index of refraction on the wave field will result in higher harmonics generation, which will eventually alter the shape of the wave pulse along the propagation.

\begin{figure}[t]
\centering
\includegraphics[width=.8\linewidth]{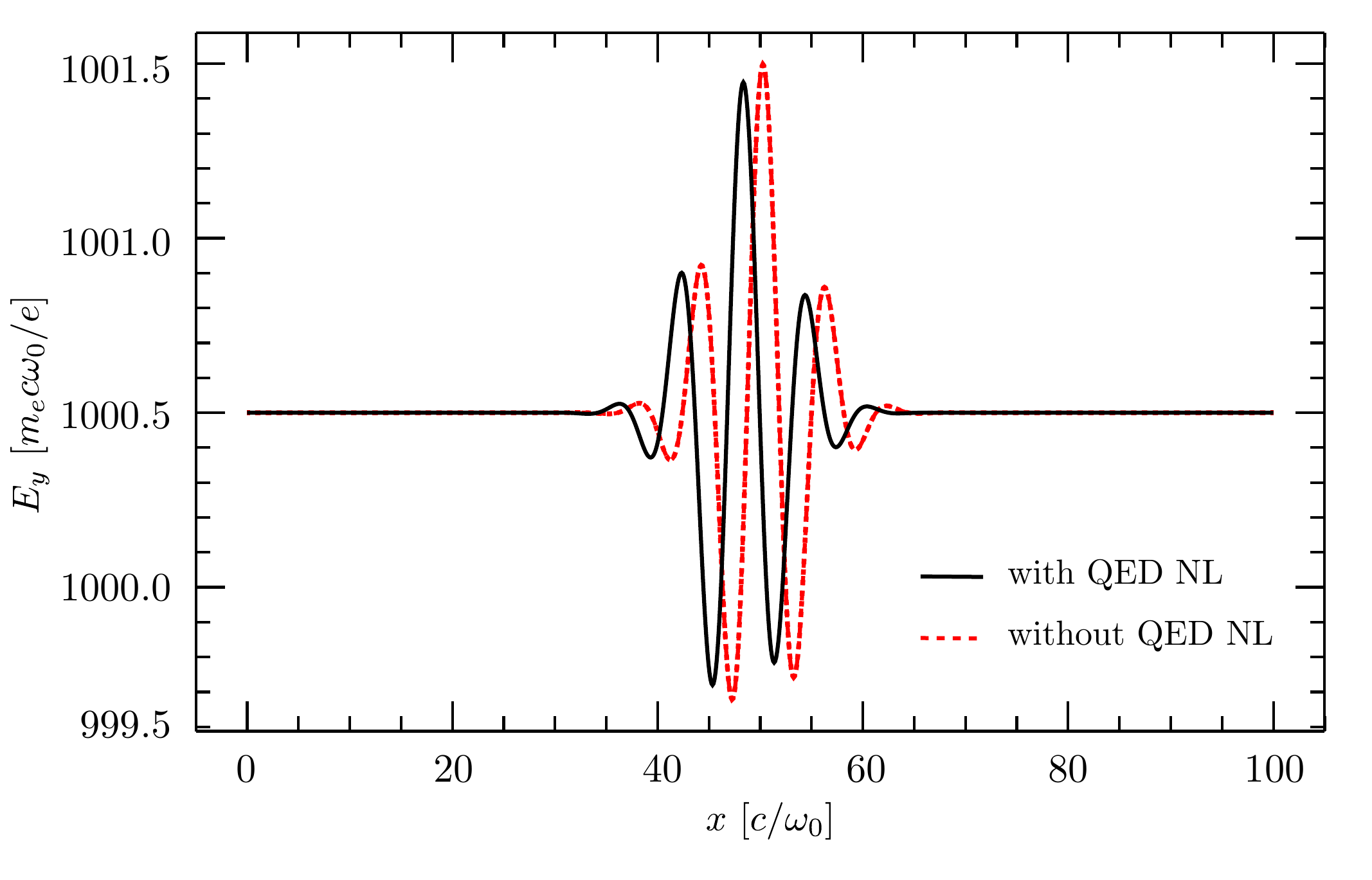}
\caption{1D Gaussian pulse after an entire propagation over a periodic box in the presence of a strong static field, with (black) and without (red) QED nonlinearities}
\label{fig:1dgaussian}
\end{figure}

\begin{figure}[t]
\centering
\includegraphics[width=.8\linewidth]{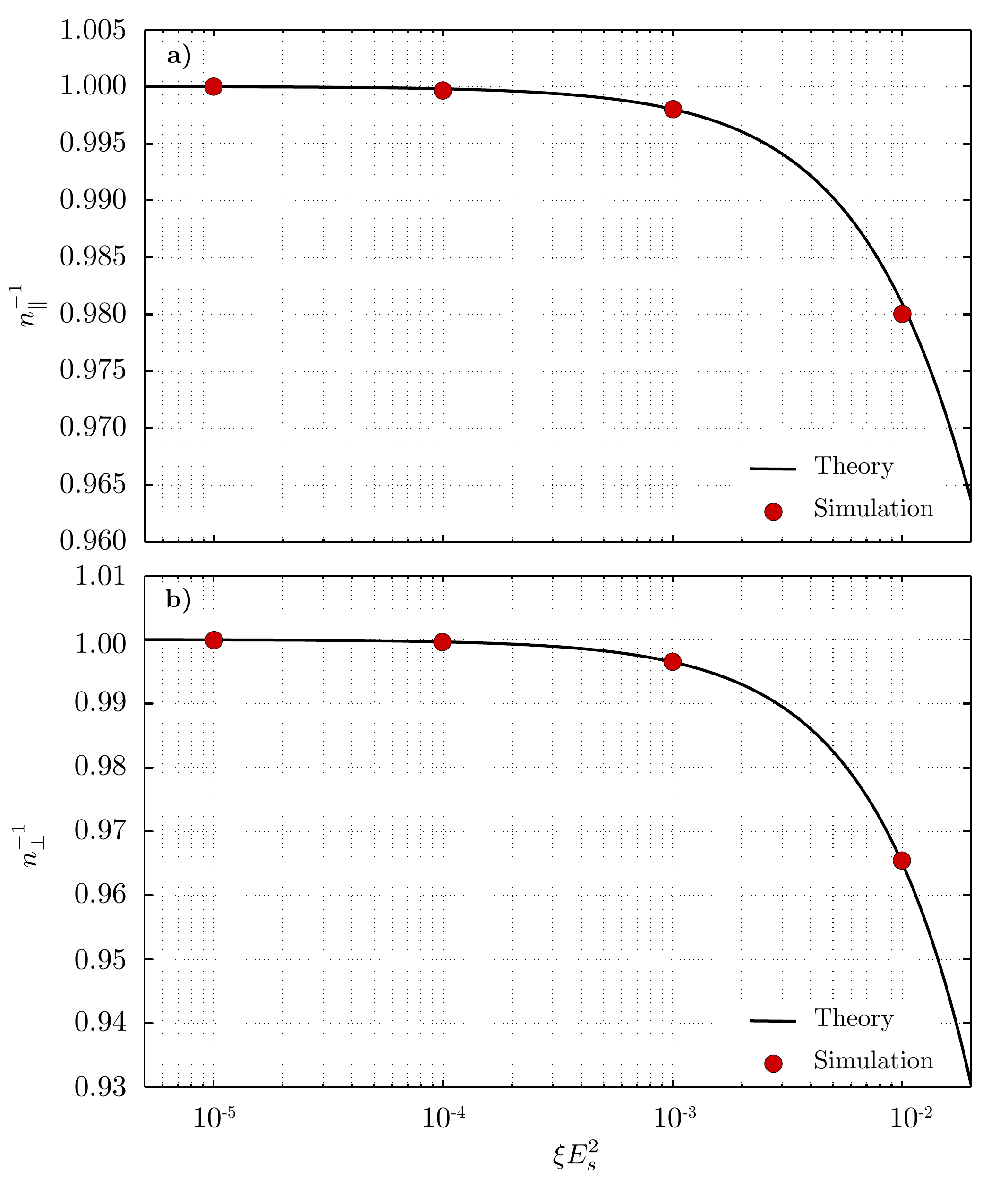}
\caption{ (a) Phase velocity $(c = 1)$ of probe pulse with  polarization parallel to $E_s$, (b) Phase velocity $(c = 1)$ of probe pulse with  polarization perpendicular to $E_s$, both as a function of $\xi E_s^2$ parameter }
\label{fig:bir}
\end{figure}

The simulation setup consists of a strong static electric field of $10^{-3}E_s$ aligned along the $y$ direction and a Gaussian EM pulse propagating in the $x$ direction and polarized in the $y-z$ plane. The central wavelength of the EM Gaussian pulse is $1 \ \mu $m and its duration is 5.6 fs. \Fref{fig:1dgaussian} shows two simulations for the same pulse after propagating once through a periodic box. In one case the propagation is in the classical vacuum, whereas the QED solver is used in the other. Qualitatively, the difference in propagation distance and the reduced electric field amplitude is consistent with the theory of a pulse traveling in a refractive medium. To test the accuracy of the algorithm this same setup was run for different values of the product $\xi E_s^2$ for both the parallel and perpendicular setup. The difference in phase velocity between the two pulses allows to extract directly the quantum vacuum refractive indices and to compare with the analytical predictions of equations
~\eref{npar} and \eref{nperp}. The results are shown in \fref{fig:bir}(a) and \fref{fig:bir}(b) where an excellent agreement between simulations and theory is found.


\subsection{Vacuum birefringence: Optical pump and X-ray probe}

The effects of the quantum vacuum on the propagation of light waves requires a very strong static electrical field that is unlikely to be produced in the laboratory. However it has been suggested a long time ago, for example by Brezin and Itzykson \cite{Brezin}, that high intensity oscillatory fields with frequency small compared with the wave frequency might play the role of an external field. This is nowadays the aim of various experiments based on the interaction between a counter-propagating ultra-intense optical pulse and a low-amplitude X-ray probe pulse \cite{King_review, HIBEF_Heinz}. It should be emphasized that vacuum birefringence differs for finite times when considering an eternally-constant background (as in section 3.1) or an adiabatically-evolved, quasi-static background~\cite{Dinu1, Dinu2}. The latter configuration has motivated the numerical work presented in this subsection

We start with the wave equation for the corrected QED Maxwell's equations \cite{DiPiazza,King1} given by:
\begin{equation}
\nabla^2 \textbf{E} -\frac{1}{c^2}\frac{\partial^2 \textbf{E}}{\partial t^2} = \textbf{S},
\label{wave}
\end{equation}
where
\begin{equation}
\textbf{S} = 4\pi \left[\frac{1}{c} \nabla\times\left(\frac{\partial \textbf{M}}{\partial t}\right)+\frac{1}{c^2}\frac{\partial^2 \textbf{P}}{\partial t^2}-\nabla\left(\nabla \cdot \textbf{P}\right)\right].
\label{source}
\end{equation}

\begin{figure}[ht]
\centering
\includegraphics[width=.8\linewidth]{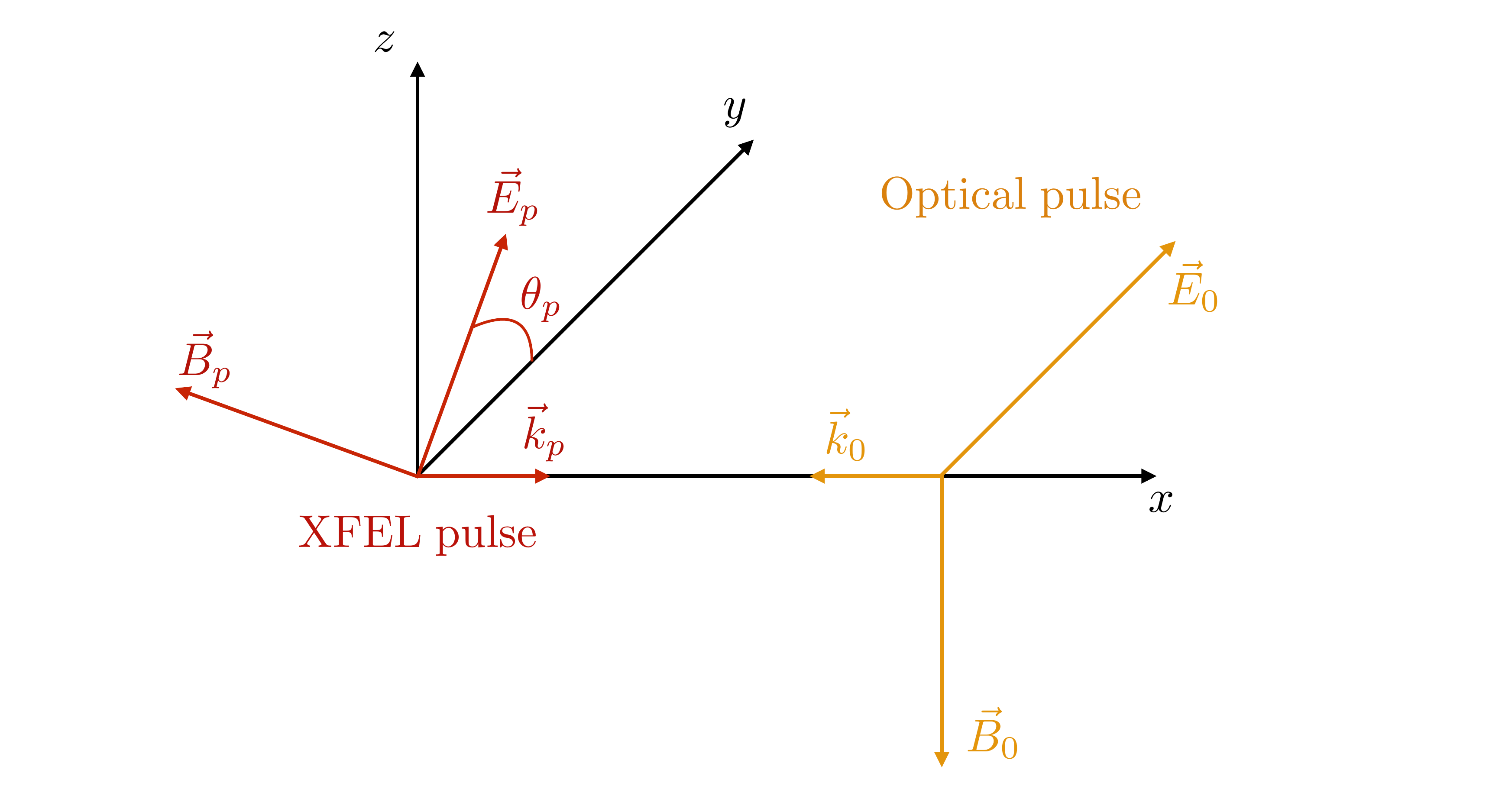}
\caption{Counter-propagating setup proposed in \cite{HIBEF_Heinz} to probe the quantum vacuum.}
\label{setup}
\end{figure}

It is evident that the probe pulse propagation will be modified whenever the source term $\textbf{S}$ is nonzero. This is verified when the EM invariants are nonzero and the strong field intensity approaches the critical value. If one assumes that the strong field remains unperturbed by the nonlinearities, then we can search for solutions of the HE Maxwell's equations of the form
\begin{equation}
\textbf{E} = \textbf{E}_p + \textbf{E}_0,
\label{E}
\end{equation}
where $p$ and $0$ stand for the probe and strong field quantities, respectively. Plugging Eq.(\ref{E}) in Eqs.(\ref{eq:P}-\ref{eq:M}) and considering just the dominant terms, we can obtain approximate expressions for the resultant probe field after the interaction. This last step can be achieved by solving equation~\eref{wave} using the Green's function method \cite{King1, DiPiazza}. The increasing availability of high-intensity lasers renders these QED effects closer to being detectable in large-scale facilities. The prospect of coupling a $1$ PW optical laser with an XFEL laser pulse will allow measuring the ellipticity induced by the QED vacuum non-linearities \cite{HIBEF_Heinz, King1}. The setup shown in \fref{setup} comprises the most promising configuration for detecting the vacuum birefringence.

For this setup, the one-dimensional approximate solutions for the probe (being a plane wave), considering just an Gaussian profile for the pump, are
\numparts
\begin{eqnarray}
&E_{py} = E_{py}^{(0)} + E_{py}^{(1)}\\
&E_{pz} = E_{pz}^{(0)} + E_{pz}^{(1)},
\end{eqnarray}
\endnumparts
where, for $\theta_p=\pi/4$,
\numparts
\begin{eqnarray}
\label{E_probe}
&E_{py}^{(0)} = E_{pz}^{(0)} = \frac{E_p}{\sqrt{2}} \cos\left(k_p\left(x-ct\right)\right)\\
&E_{py}^{(1)} = 8\sqrt{2} \pi^{3/2} \xi E_0^2 k_p  \sigma_0 \mathrm{erf}\left(\frac{x}{\sigma_0}\right) E_p \sin\left(k_p (x-ct)\right) \label{epy1}\\
&E_{pz}^{(1)} = 14\sqrt{2} \pi^{3/2} \xi E_0^2 k_p \sigma_0 \mathrm{erf}\left(\frac{x}{\sigma_0}\right) E_p \sin\left(k_p (x-ct)\right) \label{epz1},
\end{eqnarray}
\endnumparts
with $\sigma_0$ being the optical pulse length, $k_p$ the wave number of the probe, and $\theta_p$ the angle between the electric field polarization and the $y$ axis. The expressions  given by Eqs.(\ref{E_probe}) already show us several important features. The first-order probe field components have a $\pi/2$ phase-shift relative to the zeroth-order components. This makes it explicit that QED induced components are contributing to a resulting ellipsoidal polarization \cite{Dinu1,Dinu2}. The induced components scale with $\xi E_0^2$, which shows that this effect is only measurable for strong fields approaching the Schwinger critical field value. Dealing with an optical laser pulse of finite duration, the effective interaction distance is now proportional to its pulse duration parameter, $\sigma_0$. The accumulated phase after the interaction can be written as \cite{King_review}
\begin{eqnarray}
\Delta \phi &= \frac{\left|E_{pz}^{(1)}-E_{py}^{(1)}\right|}{\frac{E_p}{\sqrt{2}}\sin\left(k_p (x-ct)\right)}= 12 \pi^{3/2} \xi E_0^2 k_p \sigma_0 \mathrm{erf}\left(\frac{x}{\sigma_0}\right) \textrm{ .}
\end{eqnarray}
The ellipticity $b$ is related to $\Delta \phi$ \cite{King_review, HIBEF_Heinz}, and reads, for a static field,
\begin{equation}
b = \frac{\sin{2\theta_p}}{2}E_p\Delta \phi,
\label{ellipstatic}
\end{equation}
where $\Delta \phi = 12\pi k_pd\xi E_0^2$ and where d is the distance of interaction. For a probe interacting with an optical pump one obtains
\begin{equation}
b = 6 \pi^{3/2} E_p \sin\left(2\theta_p\right) \xi E_0^2 k_p \sigma_0 \mathrm{erf}\left(\frac{x}{\sigma_0}\right).
\label{elliplaser}
\end{equation}
The static field result is recovered if we take into account: (i) a factor of 2 that appears from considering an optical pulse with no longitudinal oscillations; (ii) the limit $\sigma_0 \rightarrow \infty$. 
Up to now, we have only considered longitudinal effects. However, finite laser pulses possess transverse Gaussian profiles. 

We study how these Gaussian profiles applied to the strong and probe pulses modify the amplitude of the ellipticity as a function of several constraints applicable to realistic configurations: probe polarization angle, laser misalignment, timing and spatial jitter. We quantify these modifications through the variation of the ellipticity due to having a more realistic optical pulse duration and pulse diffraction. The standard simulation setup is the one presented in \fref{setup}, where both laser pulses are focused at the same point in time and space and counter-propagate co-axially. The XFEL (X-ray pulse) is represented by a $3\;\mathrm{fs}$ pulse with $10^{18}\;\mathrm{W/cm^2}$ and wavelength $\lambda_p = 10\;\mathrm{nm}$. The optical pump pulse has the same duration with $10^{23}\;\mathrm{W/cm^2}$ and  $\lambda_0 = 1\;\mathrm{\mu m}$. We consider both pulses to have the same spot size $W_p = W_0 = 30\;\mathrm{[c/\omega_0]} = 4.78\;\mathrm{\mu m}$ at the focal place $x=0$. For a better illustration of the vacuum nonlinearity effect, we have increased artificially the coupling parameter to $\xi = 4\times 10^{-13}$.

\subsection{Probe polarization angle}

\begin{figure}[ht]
\centering
\includegraphics[width=.8\linewidth]{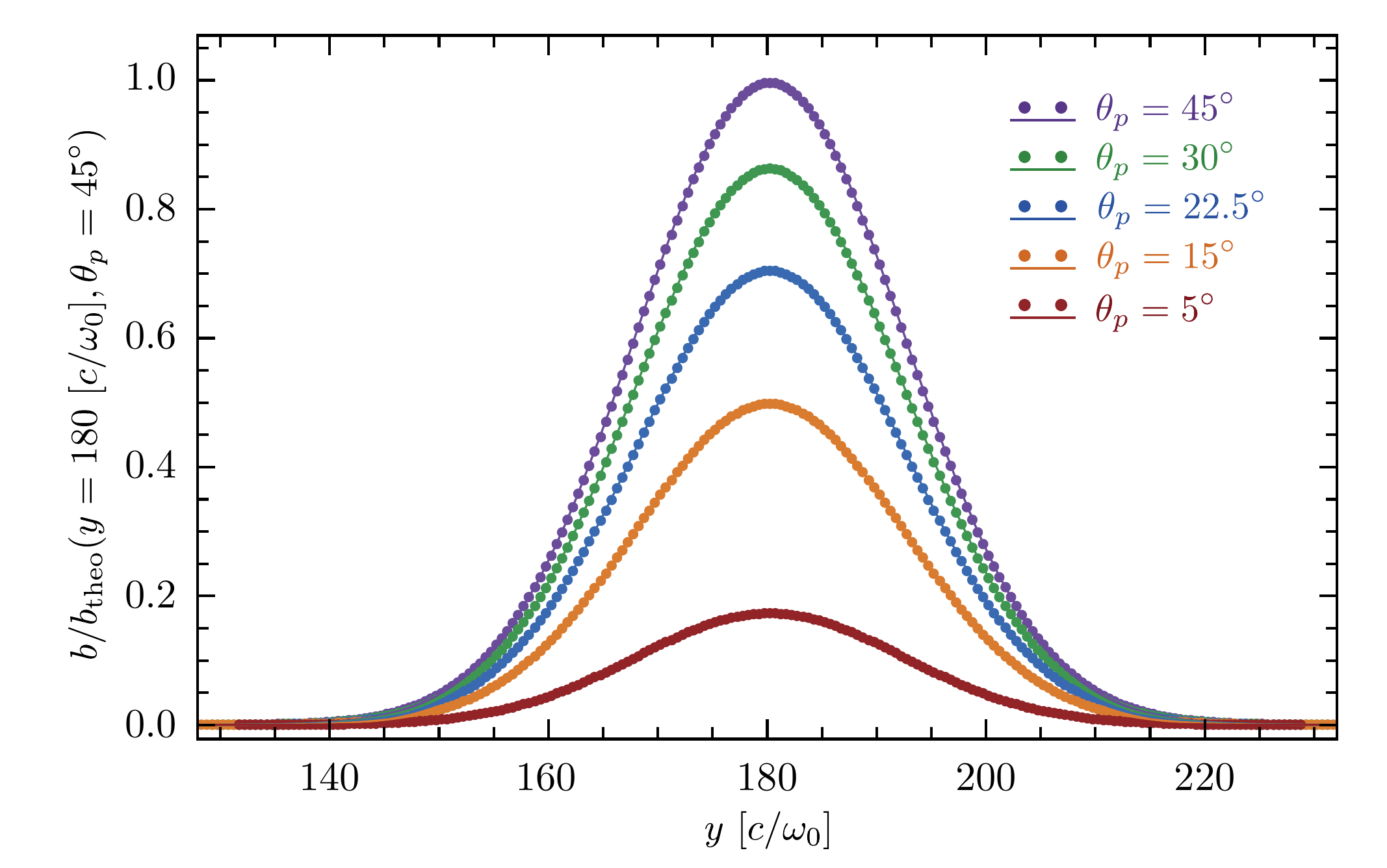}
\caption{Normalized ellipticity as a function of the transverse position for five different probe polarization angles: $45^{\circ}$ (purple), $30^{\circ}$ (green), $22.5^{\circ}$ (blue), $15^{\circ}$ (orange) and $5^{\circ}$ (brown). The colored dots and the full lines represent the simulation results and the theoretical predictions, respectively.}
\label{pol1}
\end{figure}

\begin{figure}[ht]
\centering
\includegraphics[width=.8\linewidth]{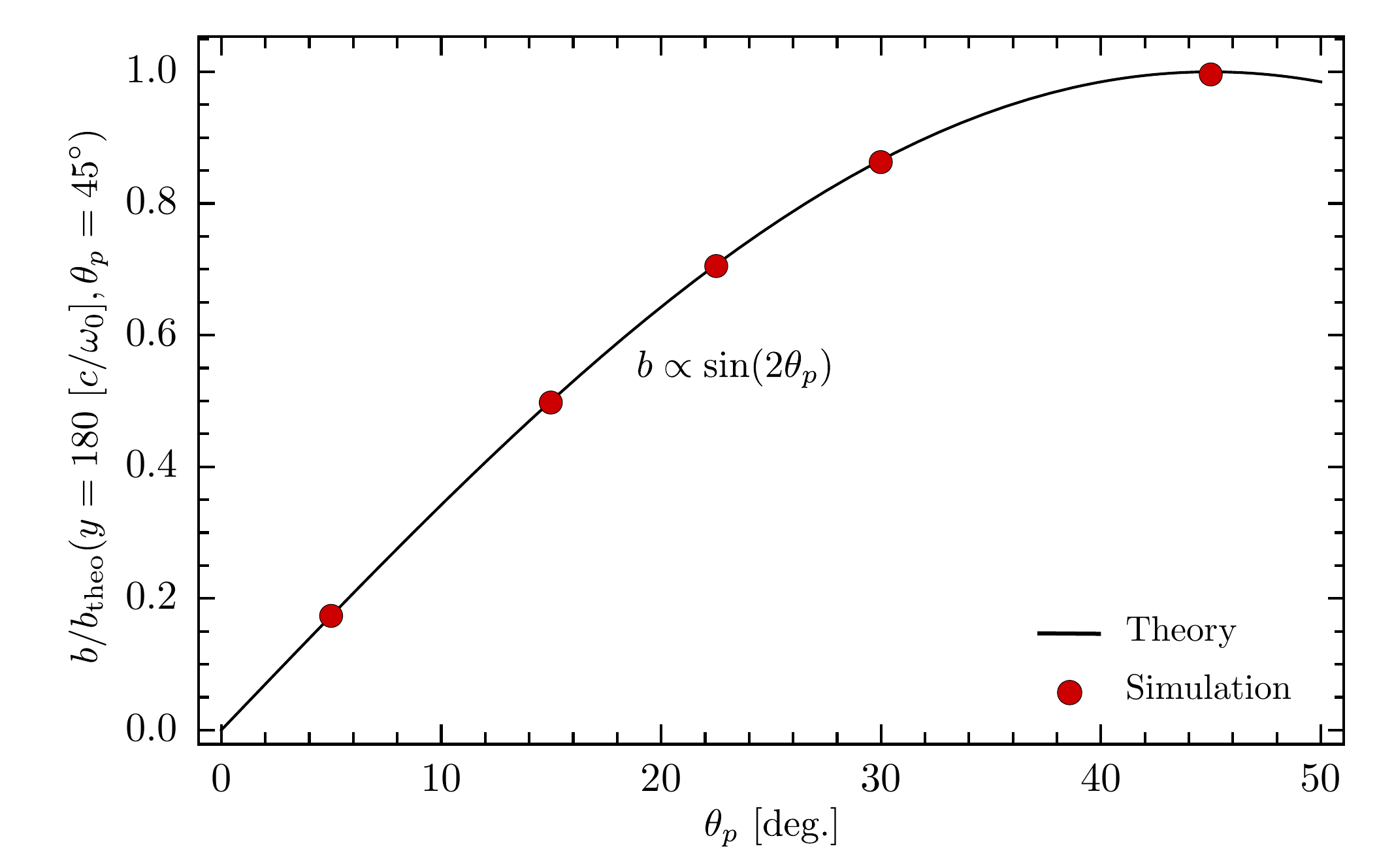}
\caption{Maximum value of the normalized ellipticity as a function of the probe polarization angle.}
\label{pol2}
\end{figure}

Equation (\ref{elliplaser}) shows that the ellipticity is proportional to $\sin\left(2\theta_p\right)$, thus yielding its maximum amplitude when the X-ray probe polarization has a $\pi/4$ angle with respect to the strong optical pulse polarization. In addition, considering a two-dimensional paraxial Gaussian transverse profile, the ellipticity will vary along the $y$ direction due to the transverse dependence of the fields
\numparts
\begin{eqnarray}
&E_p \rightarrow \frac{E_p}{\sqrt[4]{1+\left(\frac{z}{z_{\mathrm{rp}}}\right)^2}} \exp\left(- \frac{\left(y-y_p\right)^2}{W_p^2 \left(1+\left(\frac{z}{z_{\mathrm{rp}}}\right)^2\right)}\right)\label{ep} \label{eq:ep_y} \\
&E_0^2 \rightarrow \frac{E_0^2}{\sqrt{1+\left(\frac{z}{z_{\mathrm{r0}}}\right)^2}} \exp\left(- 2\frac{\left(y-y_0\right)^2}{W_0^2 \left(1+\left(\frac{z}{z_{\mathrm{r0}}}\right)^2\right)}\right)\label{e0} \label{eq:e0_y} \textrm{ ,}
\end{eqnarray}
\endnumparts
with $y_0$ and $y_p$ being the center of the Gaussian profile along $y$ for both laser pulses and $z$ being the longitudinal distance from the focus of each pulse ($z_{\mathrm{rp,0}}= k_{p,0} W_{p,0}^2/2$ is the Rayleigh length). We consider here the special case where both pulses interact in the same point in time and space ($z = 0$). \Fref{pol1} shows the transverse variation of the ellipticity for different probe polarization angles. On the other hand, the dependency of the ellipticity with the angle is verified in \fref{pol2}. The ellipticity value shown has been normalized to the best case scenario, corresponding to a transverse position $y=y_0=y_p$ and a probe polarization angle $\theta_p=\pi/4$.


\subsection{Laser misalignment}

\begin{figure}[ht]
\centering
\includegraphics[width=.8\linewidth]{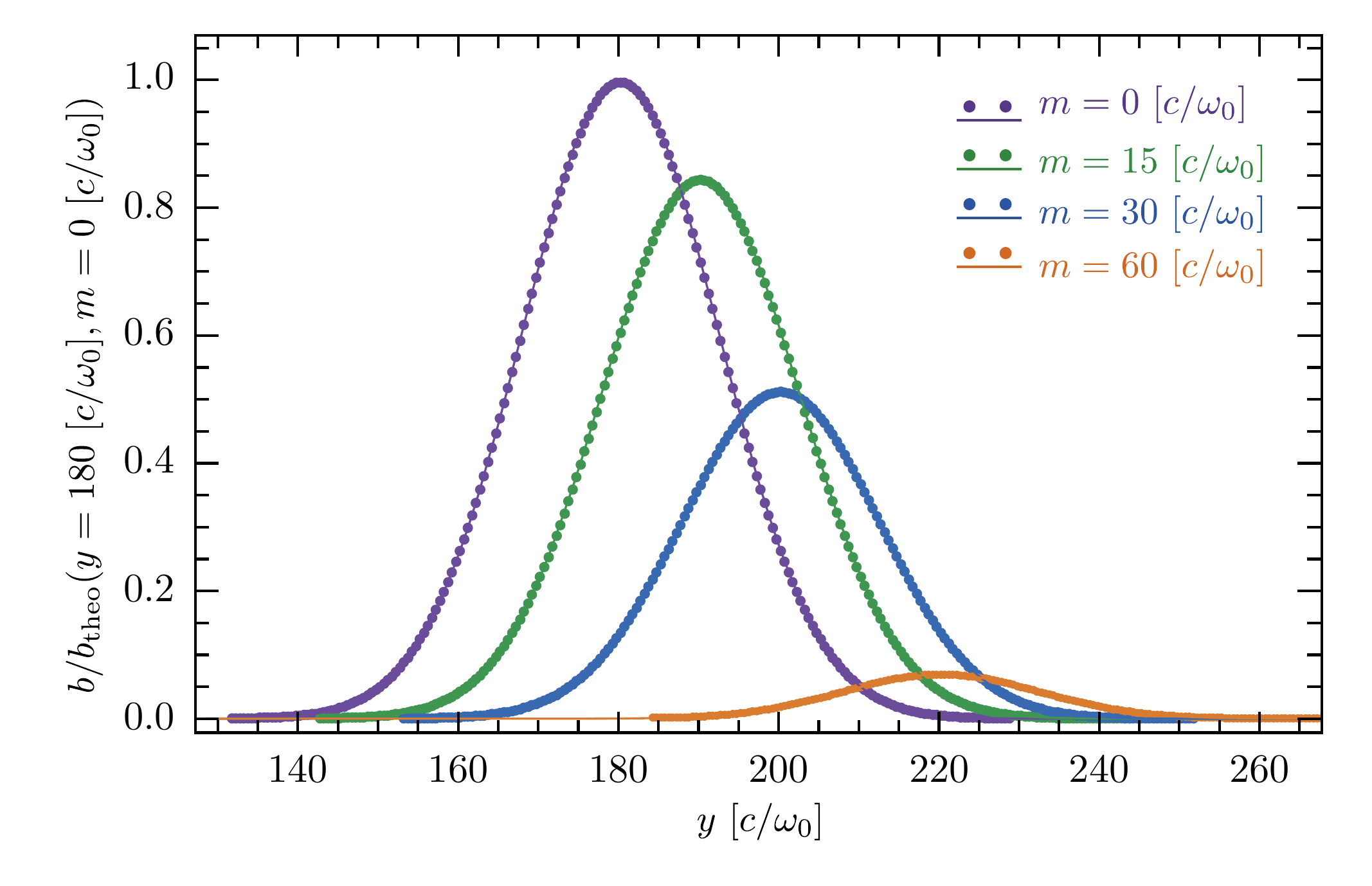}
\caption{Normalized ellipticity as a function of the transverse position for four different laser misalignment parameters: $0\;\mathrm{\mu m}$ (purple), $2.39\;\mathrm{\mu m}$ (green), $4.78\;\mathrm{\mu m}$ (blue) and $9.55\;\mathrm{\mu m}$ (orange). The coloured dots and the full lines represent the simulation results and the theoretical predictions, respectively.}
\label{mis1}
\end{figure}

\begin{figure}[ht]
\centering
\includegraphics[width=.8\linewidth]{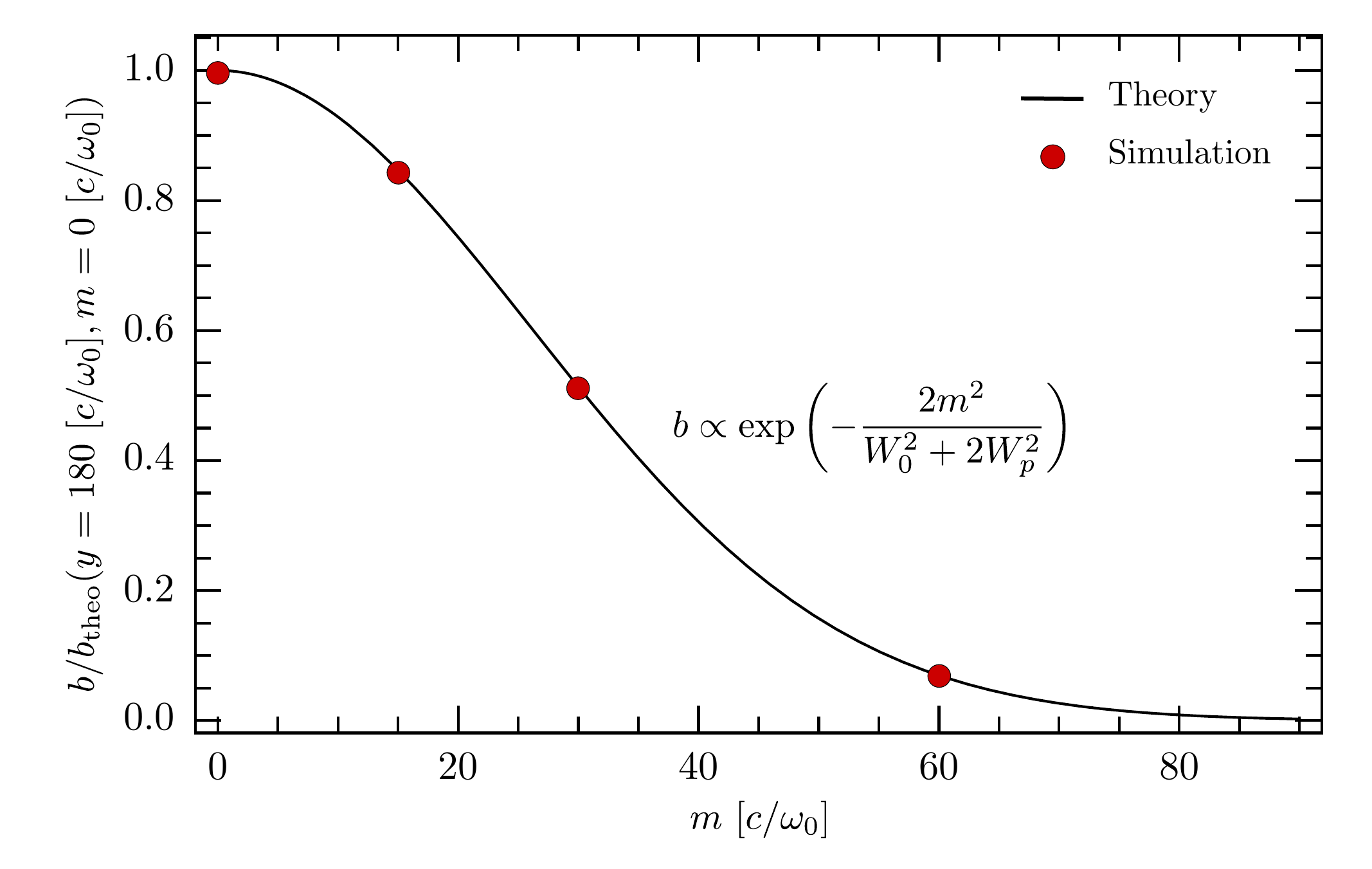}
\caption{Maximum value of the normalized ellipticity as a function of the laser misalignment parameter.}
\label{mis2}
\end{figure}

An important effect to take into account is the laser misalignment in the case where both pulses are not propagating along the same axis. In this case, the transverse profiles of the pulses do not overlap fully and we define
\begin{equation}
m=y_0-y_p
\end{equation}
as the misalignment parameter (also known as the transverse impact parameter). As equation~\eref{elliplaser} consists in the product of two transverse profiles, it is then possible to rewrite the dependency of the ellipticity on the transverse coordinate as
\begin{equation}
b(y) = \bar{b} \exp\left(-\frac{2(y-y_{\mathrm{eff}})^2}{W_{\mathrm{eff}}^2}\right)\exp(-R),\label{b}
\end{equation}
with
\numparts
\begin{eqnarray}
&\bar{b}= 6 \pi^{3/2} E_p \sin\left(2\theta_p\right) \xi E_0^2 k_p \sigma_0 \label{bbar}\\
&W_{\mathrm{eff}}^2=\frac{2W_0^2W_p^2}{W_0^2+2W_p^2}\label{Weff}\\
&y_{\mathrm{eff}}=\frac{y_pW_0^2+2y_0W_p^2}{W_0^2+2W_p^2}\\
&R=\frac{2m^2}{W_0^2+2W_p^2}. \label{R}
\end{eqnarray}
\endnumparts
It is important to notice that we have written equation~\eref{b} as an effective transverse profile, hence, the expression for $\bar{b}$ contains only the central amplitude of both pulses. \Fref{mis1} shows the transverse variation of the ellipticity for different values of misalignment, while \fref{mis2} shows how the maximum for each case vary with the misalignment parameter. As previously, the ellipticity amplitude is normalized by the best case scenario, corresponding to a coaxial counter-propagation ($m=0$) and, thus, $y=y_0=y_p$. The probe polarization angle is considered to be $\theta_p=\pi/4$. These results are consistent with the ones obtained in section III-D of \cite{Dinu1}. In this paper, the authors show that the full exponent for the ellipticity amplitude should decay with $-2\rho^2/(1+\bar{\omega})$, where $\rho=r/W_0$ and $\bar{\omega}=2W_p^2/W_0^2$. If we identify $m=r$, following the notation of \cite{Dinu1}, we verify that it is equivalent to the two-dimensional $R$ decay expression in equations~\eref{b} and \eref{R}.


\subsection{Timing and spatial jitter}

\begin{figure}[ht]
\centering
\includegraphics[width=.8\linewidth]{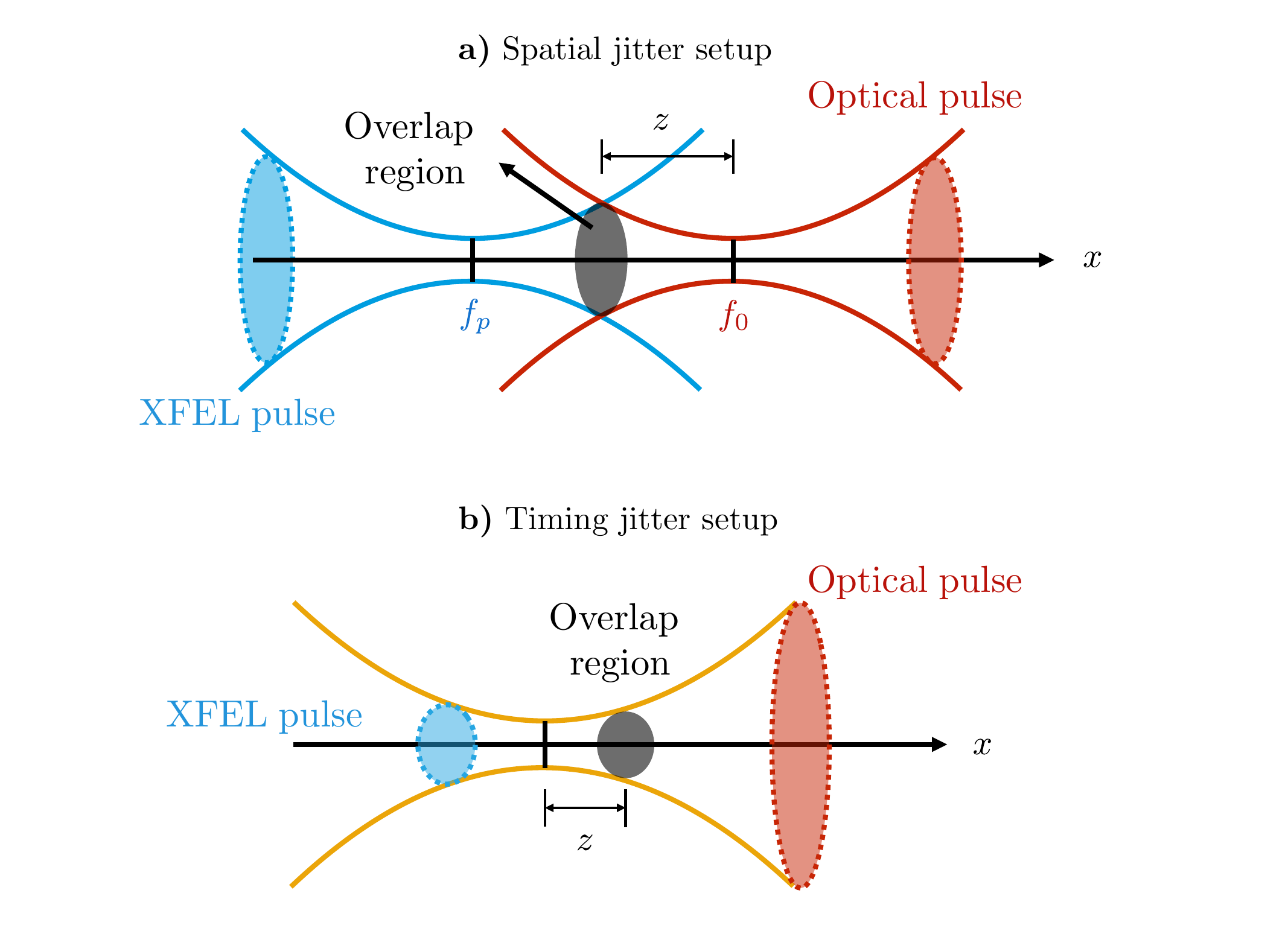}
\caption{Spatial and timing jitter setups for two counter-propagating laser pulses. $f_p$ and $f_0$ are the foci of the probe and optical pulses, respectively. $z$ corresponds to the longitudinal distance between the overlap region and the focus of the optical pulse.}
\label{jitsetup}
\end{figure}

\begin{figure}[ht]
\centering
\includegraphics[width=.8\linewidth]{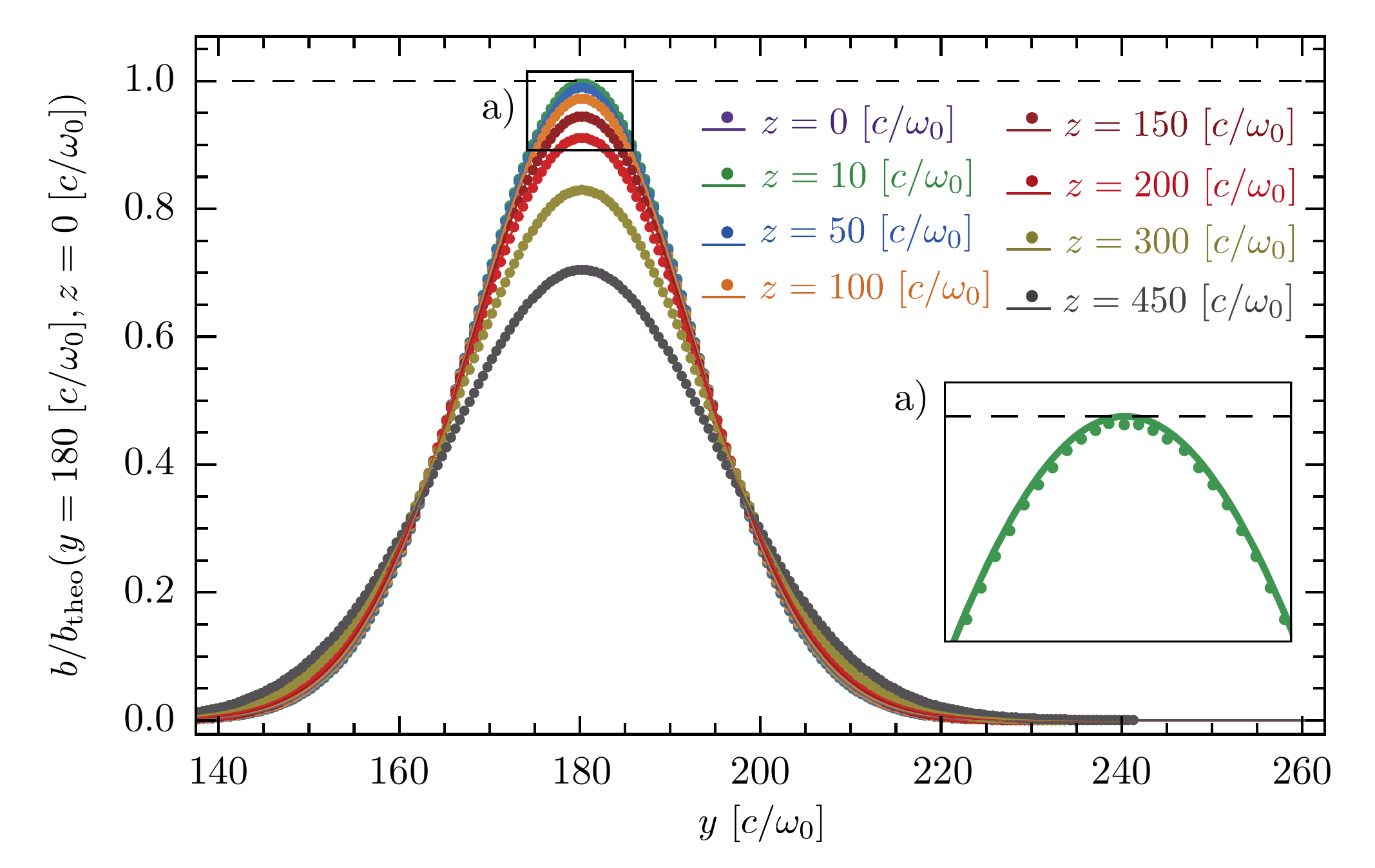}
\caption{Normalized ellipticity as a function of the transverse position for eight different values of the jitter parameter: $0\;\mathrm{\mu m}$ (purple), $1.59\;\mathrm{\mu m}$ (green), $7.96\;\mathrm{\mu m}$ (blue), $15.92\;\mathrm{\mu m}$ (orange), $23.88\;\mathrm{\mu m}$ (brown), $31.83\;\mathrm{\mu m}$ (red), $47.75\;\mathrm{\mu m}$ (greyish green) and $71.63\;\mathrm{\mu m}$ (grey, Rayleigh length). }
\label{jit1}
\end{figure}

\begin{figure}[ht]
\centering
\includegraphics[width=.8\linewidth]{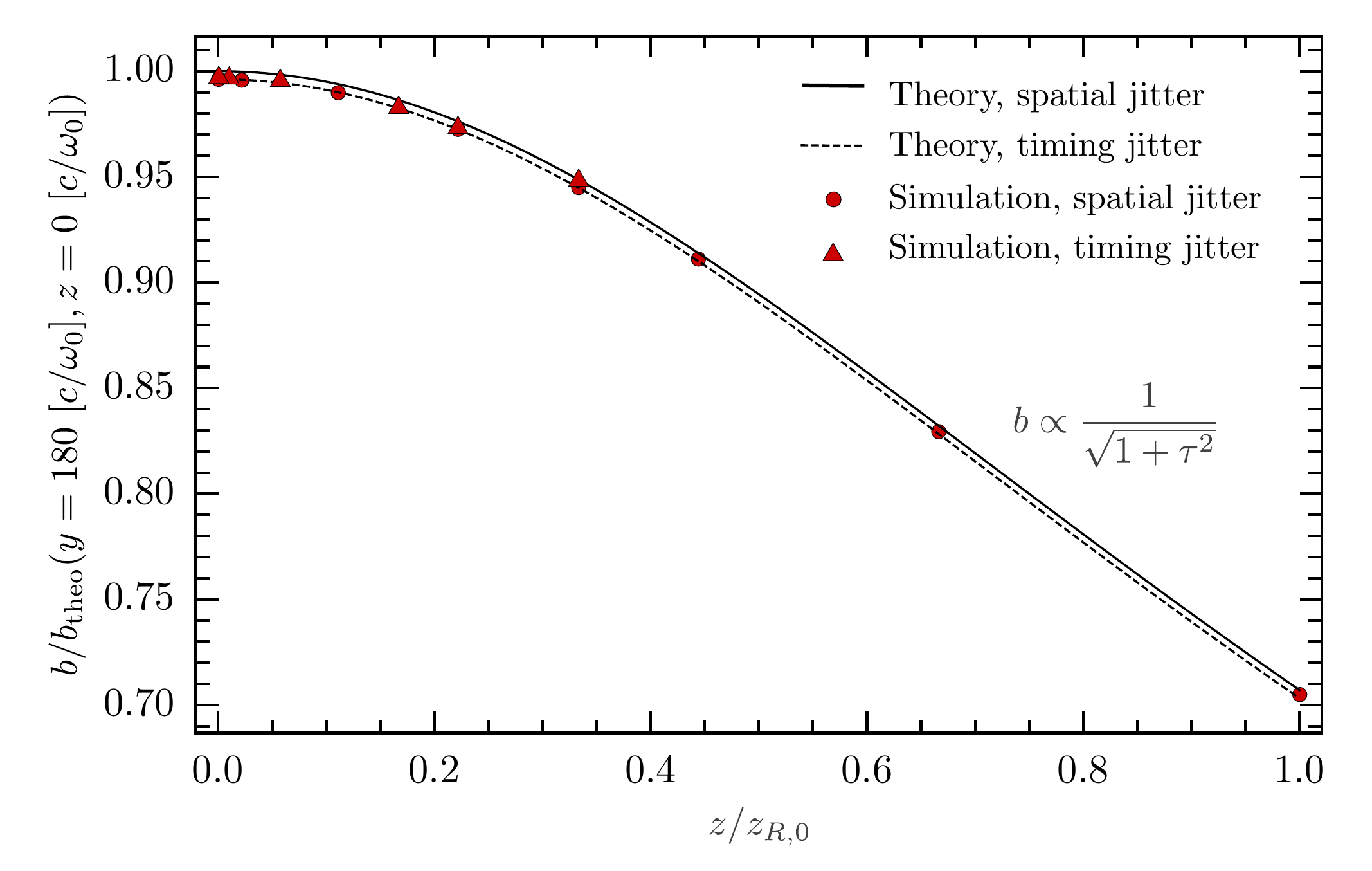}
\caption{Maximum value of the normalized ellipticity as a function of the jitter parameter.The coloured dots and the asterisks correspond to the spatial jitter and timing jitter simulation results, respectively (full or dashed lines for the theoretical predictions).}
\label{jit2}
\end{figure}

Potential fluctuations in the alignment of the lasers can lead not only to laser misalignment but also to jitter \cite{HIBEF_Heinz}. The probe pulse can be off focus with the optical pulse both in space (spatial jitter) and in time (timing jitter), as shown in \fref{jitsetup}. This effect has a direct impact on the integrated amplitude of the optical pulse as seen by the XFEL probe pulse during the overlap phase. Consequently, the ellipticity can be written as in equations~\eref{b}-\eref{R}, with $W_0^2$ replaced by $W_0^2\left(1+\left(z/z_{r0}\right)^2\right)$.
The small corrections were not taken into account in equation~\eref{eq:ep_y} since the XFEL Rayleigh length is much larger than the interaction length (approximately the optical pulse width). \Fref{jit1} shows the transverse variation of the ellipticity for different spatial jitter configurations, while \fref{jit2} shows how the maxima for each case vary with the distance to the focus of the optical pulse. In this last figure, we also superimpose results for timing jitter configurations, verifying that the results follow the same trend. The ellipticity amplitude is as usual normalized by the best case scenario, corresponding to a coaxial counter-propagation ($m=0$), probe polarization angle $\theta_p=\pi/4$, and nonexistent jitter ($z=0$).

In section III-C of \cite{Dinu2}, the authors explore these effects in a three-dimensional configuration. This implies that the Gaussian paraxial fields will decay faster, hence, yielding that the ellipticity should decay with $1/(1+\tau^2)$, where $\tau=z/z_{r0}$, instead of our square root decay. Our results and our numerical solutions are consistent with the predictions of \cite{Dinu2} further validating our algorithm.

\section{High Harmonic Generation}
\label{sec:4}

\subsection{Counter-propagating plane waves}

\begin{figure}
\centering
\includegraphics[width=.8\linewidth]{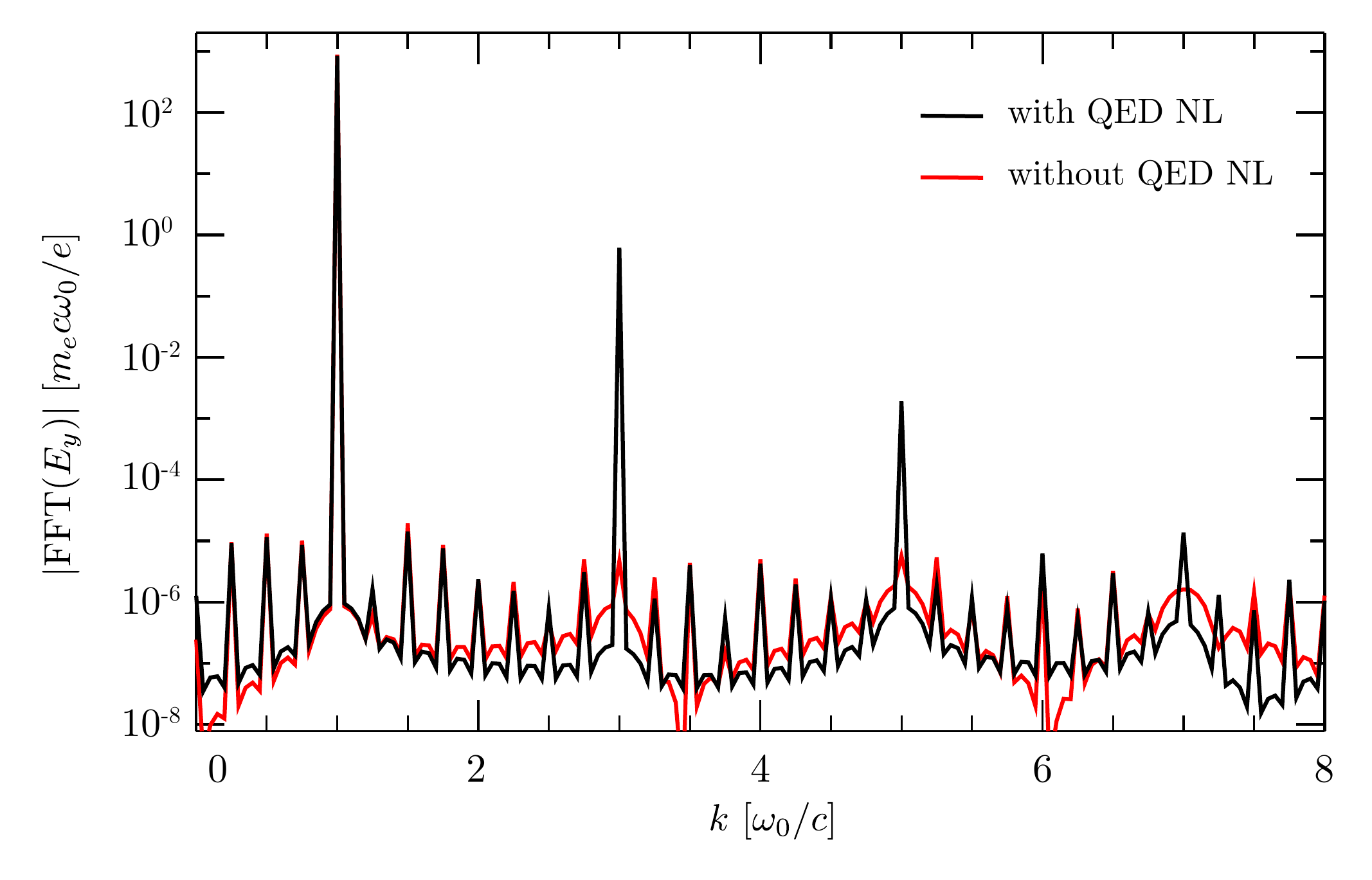}
\caption{Spatial Fourier transform of electric field with and without QED NL present.  The generation of odd higher harmonics can be observed in blue. }
\label{fig:counter_1d_fft}
\end{figure}
\begin{figure}
\centering
\includegraphics[width=.8\linewidth]{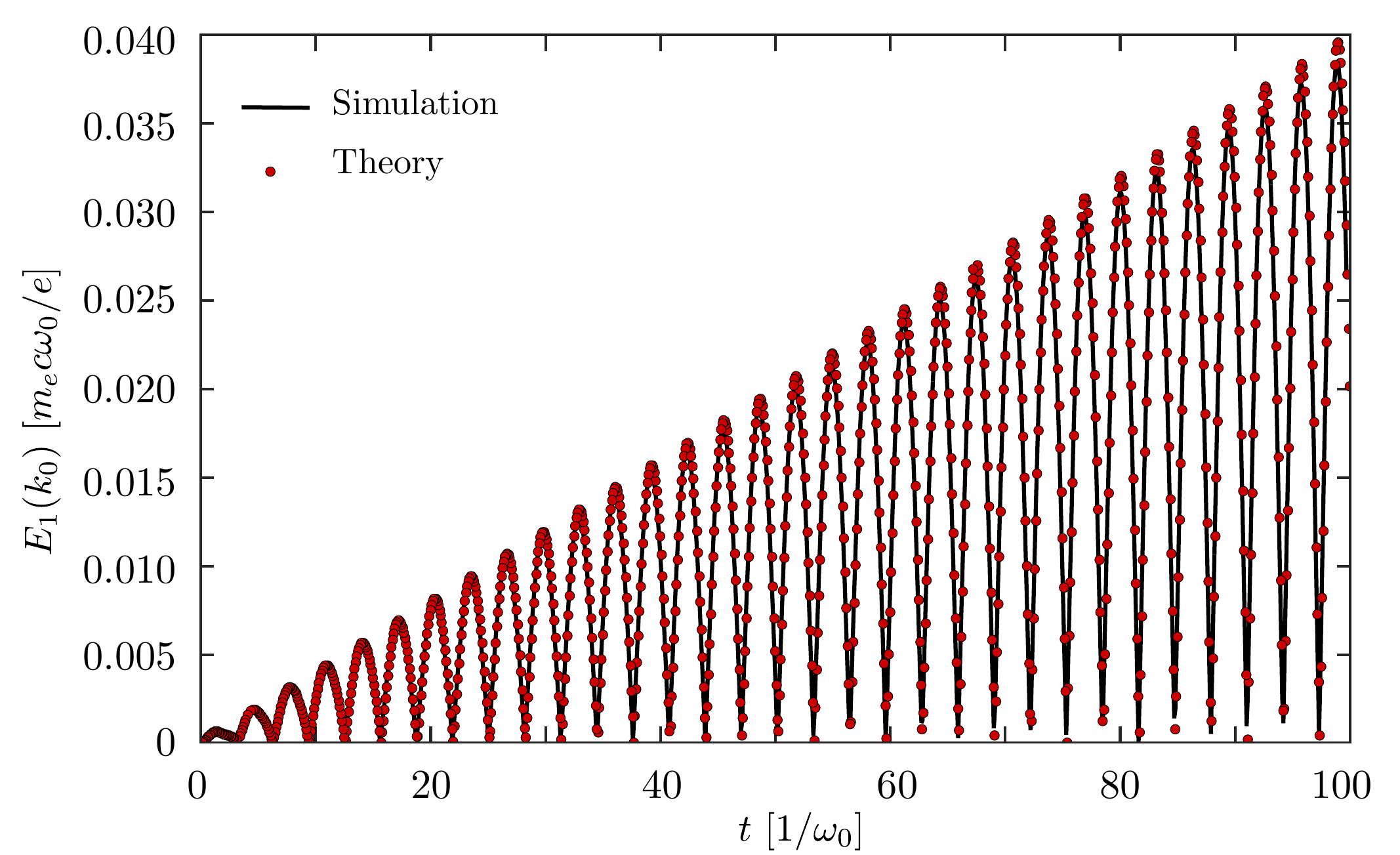}
\caption{Temporal evolution of $k_0$ Fourier mode of the subtracted electric field }
\label{fig:temporal_fft_comparison}
\end{figure}

Two counter-propagating plane waves polarized in the same direction, with the same frequency $\omega_0$ and amplitude would normally result in an electromagnetic standing wave in the classical vacuum. However, the vacuum nonlinearities lead to the well-known phenomena of harmonic generation \cite{Dipiazza_HHG, King-Keitel}. This simple setup represents an ideal benchmark for the numerical algorithm as analytic results can be obtained. To address the problem, it is convenient to decompose the field into a Born series of partial waves as performed by Bohl {\it et al.} \cite{King1}
\begin{eqnarray}
\label{expansion_1}
E &=& E_{(0)} + E_{(1)} + E_{(2)} + ... \\
\label{expansion_2}
B &=& B_{(0)} + B_{(1)} + B_{(2)} + ... ~~,
\end{eqnarray}
where $E_{(0)}$ and $B_{(0)}$ are the unperturbed standing wave fields given by $E_{(0)} = E_0 \left[ \cos(x-t) + \cos(x+t) \right]$ and $B_{(0)} = B_0 \left[\cos(x+t) - \cos(x-t) \right]$ whilst the remaining terms are successively higher order corrections to the standing wave fields, weighted by an expansion parameter to be identified. In this subsection, space and time are taken in units of $k_0^{-1}$, and $\omega_0^{-1}$ with $\omega_0 = k_0 c$. The electric and magnetic fields are aligned with the $y$ and $z$ axis, respectively. Starting from the modified Maxwell's equations and inserting Eqs.(\ref{expansion_1}-\ref{expansion_2}) as the expressions for the fields, we arrive at the wave equation for the first-order correction to the electric field $E_{(1)}$
\begin{equation}
\Box \ E_(1) = S_1(x,t),
\end{equation}
where $\Box$ is the d'Alembert operator and the source term $S_1 = - \partial_t  \partial _x {M} + \partial_t^2 {P}$. These are just the one-dimensional versiosn of equations~\eref{wave}-\eref{source}. Inserting the zero order field in the source term, i.e. taking $P, M = f({E}_{(0)},{B}_{(0)})$, we find
\begin{equation}
S_{(1)}(x,t)= 16  \xi E_0^3 \cos(t)\cos(x)
\times\left[3\cos(2t)-\cos(2x)\right].
\end{equation} 
This source term only accounts for the unperturbed fields being inserted into the nonlinear polarization and magnetization. The formal solution of this equation is given by the convolution between the source term and the  Green's function of the one-dimensional wave operator~\cite{Jackson},
\begin{equation}
E_{(1)}(x,t)=\int_{0}^{L}\int_{0}^{t} dt'dx' G(x,x',t,t')S(x',t')
\end{equation}
where the Green's function is  
\begin{equation}
G(x, x', t, t') = \frac{1}{2} H\left[(t-t')- |x-x'|\right].
\end{equation}
The modified electric field reads
\begin{equation}
E_{(1)}(x,t)= -2\xi E_0^3\sin(t) \cos(x)  
\times \left[2 \sin(2 t) \left(\cos(2 x)-2\right)-4t\right]. 
\end{equation}
We notice that the relative amplitude between $E_{(1)}$ and the unperturbed field amplitude is proportional to $\xi E_0^2$ showing again that this perturbative treatment is valid as long as $\xi E_0^2 \ll 1$. More specifically, the corrected field exhibits a secular growth term modulated by an oscillating term. This term is dominant for $t \gg 1$ and should be interpreted as a phase shift due to the induced birefringence of one wave to the other. The total field for $t \gg 1$ reads $E \simeq E_{(0)} + 8\xi E_0^3t\sin(t)\cos(x)$. Using the trigonometric identity $a\cos(x) + b\sin(x) = \sqrt{a^2+b^2}\cos(x-\arctan(b/a))$, one can write the total corrected field as 
\begin{eqnarray}
E &\simeq& E_0\left[ \cos(x-t/n) + \cos(x+t/n) \right],
\end{eqnarray}
where $n^{-1} = 1 + 4 \xi E_0^2$ and $n$ the modified refractive index induced by the interaction of the two waves. Taking the spatial Fourier transform of $E_{(1)}$, we verify that the fundamental mode $k = k_0$ is corrected by a secular term and the appearance of an harmonic at $3k_0$. Defining the Fourier transform of $E(x,t)$ as $\tilde E(k,t)$, we obtain
\begin{eqnarray}
\label{eq:eko}
\tilde E_{(1)}(k_0) &=& \xi E_0^3\sin(t)\left[4t + 3\sin(2t)\right], \\
\label{eq:e3ko}
\tilde E_{(1)}(3k_0) &=& -\xi E_0^3\sin(t)\sin(2t).
\end{eqnarray}
The third harmonics correspond to two waves, $\cos[3(x+t)]$ and $\cos[3(x-t)]$, propagating with the initial zeroth-order field. The next order correction to the field $E_{(2)}$, reveals a correction to the $k_0$ mode growing as $t^2$, a secular $3k_0$ harmonic and an oscillating $5k_0$ harmonic.  Repeating this process to higher orders, we can show that this nonlinear interaction generates odd higher harmonics from vacuum with the relative amplitude between these harmonics obeying the ordering 
\begin{equation}
\tilde{E}(k = 2n+1) = (\xi E_0^2)^n \tilde{E}(k_0).
\label{eq:ordering}
\end{equation}
Nonetheless, it should be emphasized that a rigorous treatment of higher harmonics (beyond the first-order correction) should take into account additional terms in the expansion of the Heisenberg and Euler Lagrangian for weak fields ($E < E_{sch}$). As shown by Bohl {\it et al.}~\cite{King1}, purely four-photon scattering (first-order term of the Euler Heisenberg Lagrangian) allows the generation of higher harmonics in the counter propagating setup. However the contribution from this twice-iterated process scales as $(\xi E_0^2)^2E_0$, which when compared to the leading contribution to the fifth harmonic from six photon scattering is suppressed by a factor of $\xi E_{0}^2$.

Our analytical predictions were compared with the results of the QED solver using a field amplitude of $E_0 = 0.025 E_{sch}$, $\lambda_0 = 1 \ \mu $m plane waves and $\xi =  10^{-9} $, such that the higher harmonics can be accurately resolved above the numerical noise. The spatial Fourier transform of the fields is shown in \fref{fig:counter_1d_fft} for two simulations, with and without the self-consistent inclusion of Heisenberg-Euler corrections. We observe that when the non-linearities are present, the odd higher harmonics are generated with a relative amplitude that matches the ordering given in equation~\eref{eq:ordering}. To compare the simulation results with equations~\eref{eq:eko}-\eref{eq:e3ko}, we subtracted the classical vacuum electric field to remove the zeroth order standing wave contribution, and performed the Fourier transform of this subtracted field. Finally, we tracked the temporal evolution of the amplitude of the $k_0$ mode in Fourier space and compared it with equation~\eref{eq:eko}. \Fref{fig:temporal_fft_comparison} shows the temporal evolution of $E_1(k_0)$. The simulation shows an excellent agreement with the theoretical predictions for many laser cycles, ensuring that the algorithm is robust. Despite the one-dimensionality of this example, a setup of counter-propagating beams is of great interest for planned experiments at extreme high intensity laser facilities, as outlined in~\cite{DiPiazza}. 

\subsection{Interaction of paraxial beams: counter propagating setup}

\begin{figure}[t]
\centering
\includegraphics[width=.8\linewidth]{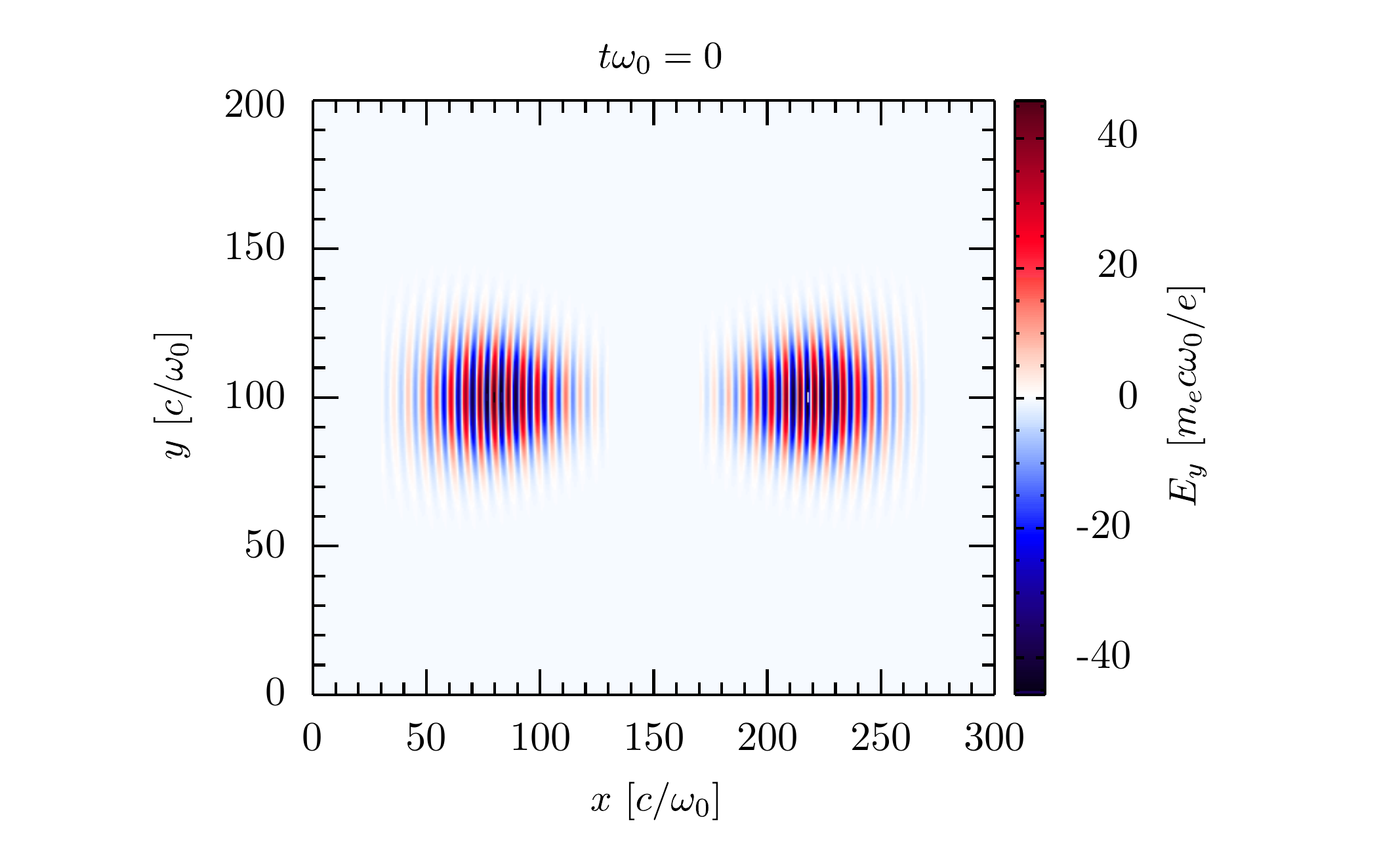}
\caption{Initial setup of Gaussian pulses. Both pulses are polarized in the x$_2$ direction and will focus in the center of the box.}
\label{fig:counter_2d_setup}
\end{figure}
\begin{figure}
\centering
\includegraphics[width=.8\linewidth]{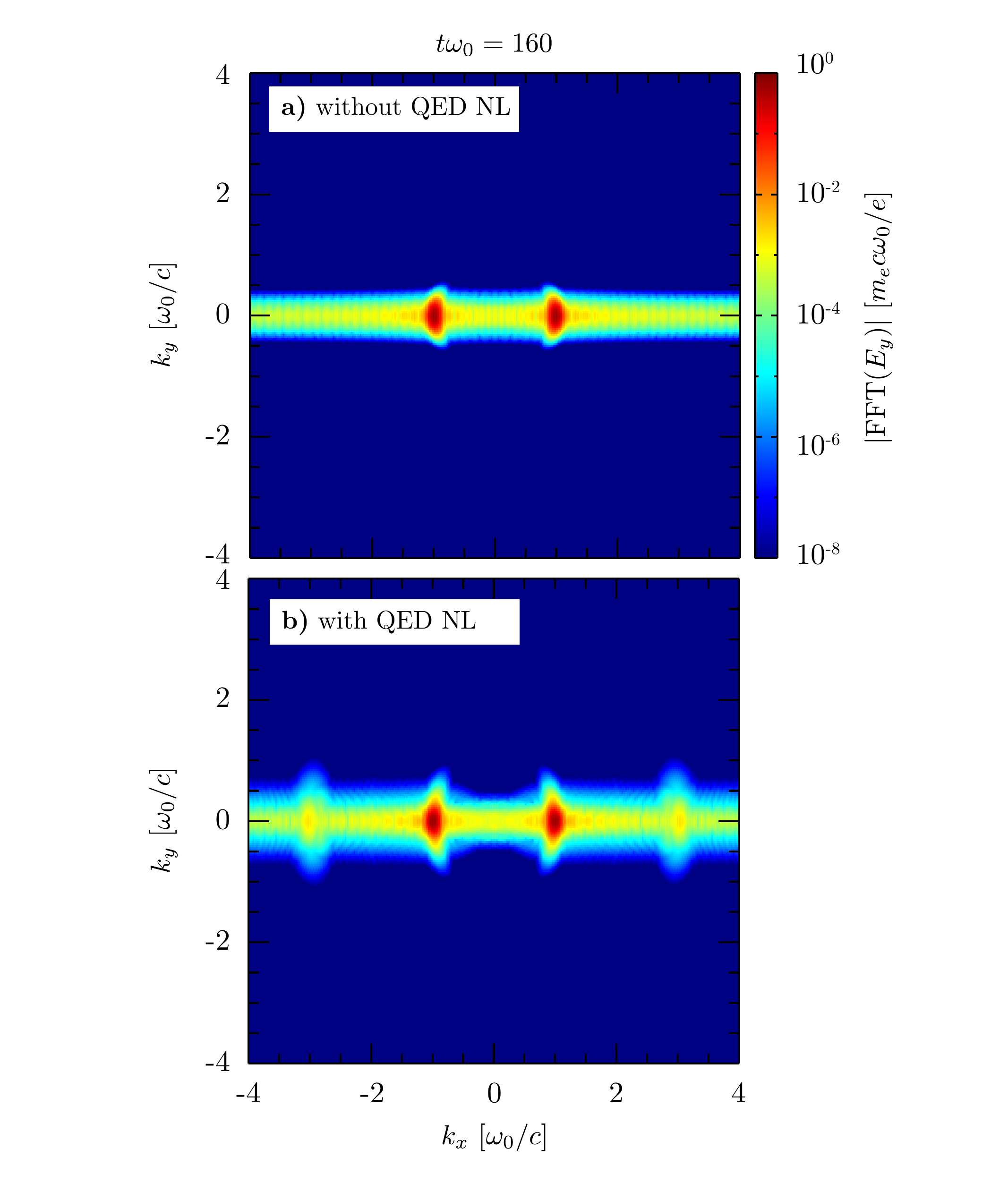}
\caption{Spatial Fourier transform of the electric field, (a) after the interaction but when QED corrections are absent, (b) after the interaction with self-consistent inclusion of the quantum corrections. The third harmonic and small distortion of the main mode can be observed.}
\label{fig:counter_2d_art}
\end{figure}

In order to illustrate the generation of harmonics in multi-dimensions, two setups were investigated: the counter propagation of two Gaussian pulses interacting at the focal point, and the perpendicular interaction of two Gaussian pulses focused in the same region of space. For these setups, a consistent analytical treatment becomes cumbersome especially due to the self-consistent treatment of both the transverse and longitudinal components of the pulses. A quantum parameter of $\xi =  10^{-7}$ was used for the sake of showing the prominent features of the harmonics of very small amplitudes. 

In the first setup, two $\lambda = 1~\mu$m laser beams with a normalized vector potential $a_0 = eE_0/mc\omega_0 = 50$ (which corresponds to laser electric field at the focus of $E_0\simeq 10^{-4}E_s$) and duration of 25 fs interacted in the presence of vacuum non-linearities. Both beams have a focal spot of $W_0 = 4~\mu$m. \Fref{fig:counter_2d_setup} shows the transverse electric field of the laser beams before interaction at $t=0$, \fref{fig:counter_2d_art}-(a) the spatial Fourier transform of the beams with $\xi = 0$ (classical limit) and in \fref{fig:counter_2d_art}-(b) the Fourier transform of the electric field after the interaction (asymptotic state) including the HE corrections. As shown in \fref{fig:counter_2d_art}-(b), after the interaction odd higher harmonics are also generated as in the 1D case, with relative amplitudes consistent with Eq.(\ref{eq:ordering}). However, in this case, the harmonics generated have the same Gaussian behavior as the unperturbed pulses and attain a greater spread in Fourier space after the interaction. After the pulses have spatially overlapped, the harmonics propagate and leave an imprint of the nonlinear interaction, that co-propagates with the original beams. 
\begin{figure}[t!]
\centering
\includegraphics[width=.8\linewidth]{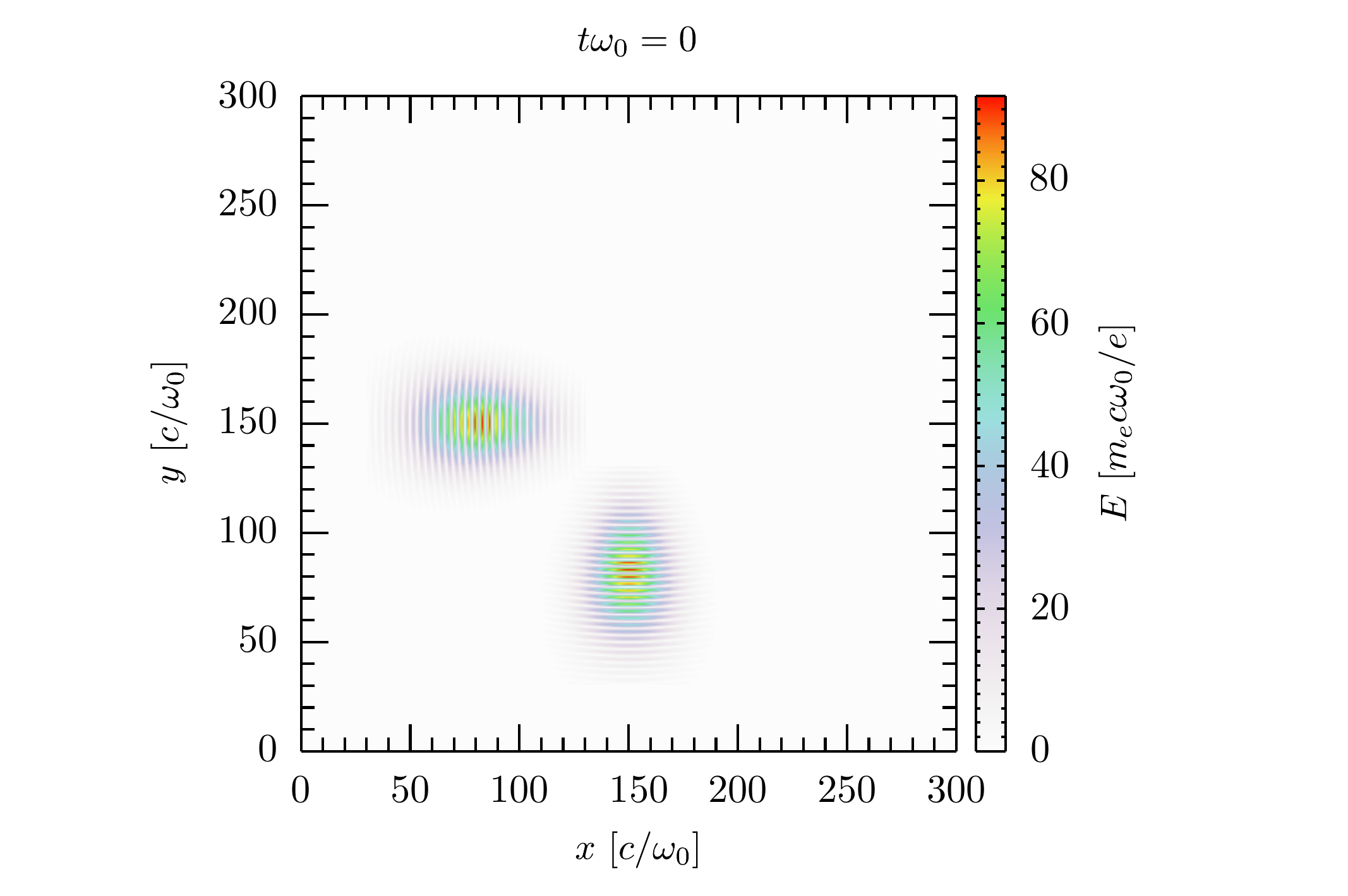}
\caption{Electric field setup for two Gaussian pulsed travelling in perpendicular directions but focusing on the centre of the simulation box.}
\label{fig:counter_prop_setup}
\end{figure}

\subsection{Interaction of paraxial beams: 90$^\circ$ setup}

The second setup, shown in Fig \ref{fig:counter_prop_setup}, comprises two optical Gaussian pulses interacting at 90$^\circ$.  The two laser beams possess the same parameters: $a_0$ = 100, a wavelength of $1~\mu m$, a focal spot $W_0 = 4 \mu  $m, a duration of 25 fs. The advantage of this setup is the vast amount of harmonics generated during the interaction of the two pulses. Before discussing the results of the simulations, the reader can develop a valuable intuition of the generated harmonics by carefully computing the electromagnetic invariants and the associated vacuum polarizations for paraxial beams.

The theory of paraxial electromagnetic fields has been developed by Davis \cite{Davis} with a simple method that allows to find a formal solution of a light beam propagating in classical vacuum. The formal solution is based on an expansion in powers of a small parameter $s = W_0/l_r = 1/kW_0$ where $W_0$ is the beam waist and $l_r = kW_0^2$ the Rayleigh or diffraction length. For the special case of a two-dimensional beam, varying as $e^{i\omega_0 t}$ propagating in the $x$ direction and polarized in the $y$ direction, the non vanishing components up to the second order in $s$ are 
\begin{eqnarray}
\label{eq:beam_davis1}
\nonumber
&E_y& = \bar{E}_y~e^{-ik_0x} + c.c. \\
&E_x& = \bar{E}_x~e^{-ik_0x} + c.c. \\
\nonumber
&B_z& = \bar{B}_z~e^{-ik_0x} + c.c,
\end{eqnarray}
with
\begin{eqnarray}
\label{eq:beam_davis2}
\nonumber
&\bar{E}_y& = -ik_0\left[\Psi_0 + s^2\left(\Psi_2 + \frac{\partial^2\Psi_0}{\partial\eta^2}\right)\right]  \\
&\bar{E}_x& = -k_0s\frac{\partial\Psi_0}{\partial\eta}  \\
\nonumber
&\bar{B}_z& = -ik_0\left[\Psi_0 + s^2\left(\Psi_2 + \frac{\partial\Psi_0}{\partial\zeta}\right)\right],
\end{eqnarray}
where $\omega_0 = k_0c$, $x = \zeta l_r$, $y = W_0\eta$ and 
\numparts
\begin{eqnarray}
\Psi_0 &=& A_0e^{-(t-x/c)^2/2\sigma^2}e^{-i(P+Q\eta^2)} \\
\Psi_2 &=& iQ(2 + Q^2\eta^4)\Psi_0 \\
Q &=& \frac{1}{i+2\eta} \\
iP &=& -\log(iQ)
\end{eqnarray}
\endnumparts
where $\sigma$ is the typical duration of the beam and $a_0 = eA_0/mc^2$ is the Lorentz invariant parameter which measures the magnitude of the field.
Whereas Gaussian paraxial beams are usually described up to the first order in $s$ \cite{Gies,King-Keitel}, the inclusion of the second-order terms for the transverse components are essential to calculate accurately the electromagnetic invariants which consist of a series of terms proportional to $(k_0A_0)^2$, $s(k_0A_0)^2$ and $(sk_0A_0)^2$ .
\begin{figure}[ht]
\centering
\includegraphics[width=0.8\linewidth]{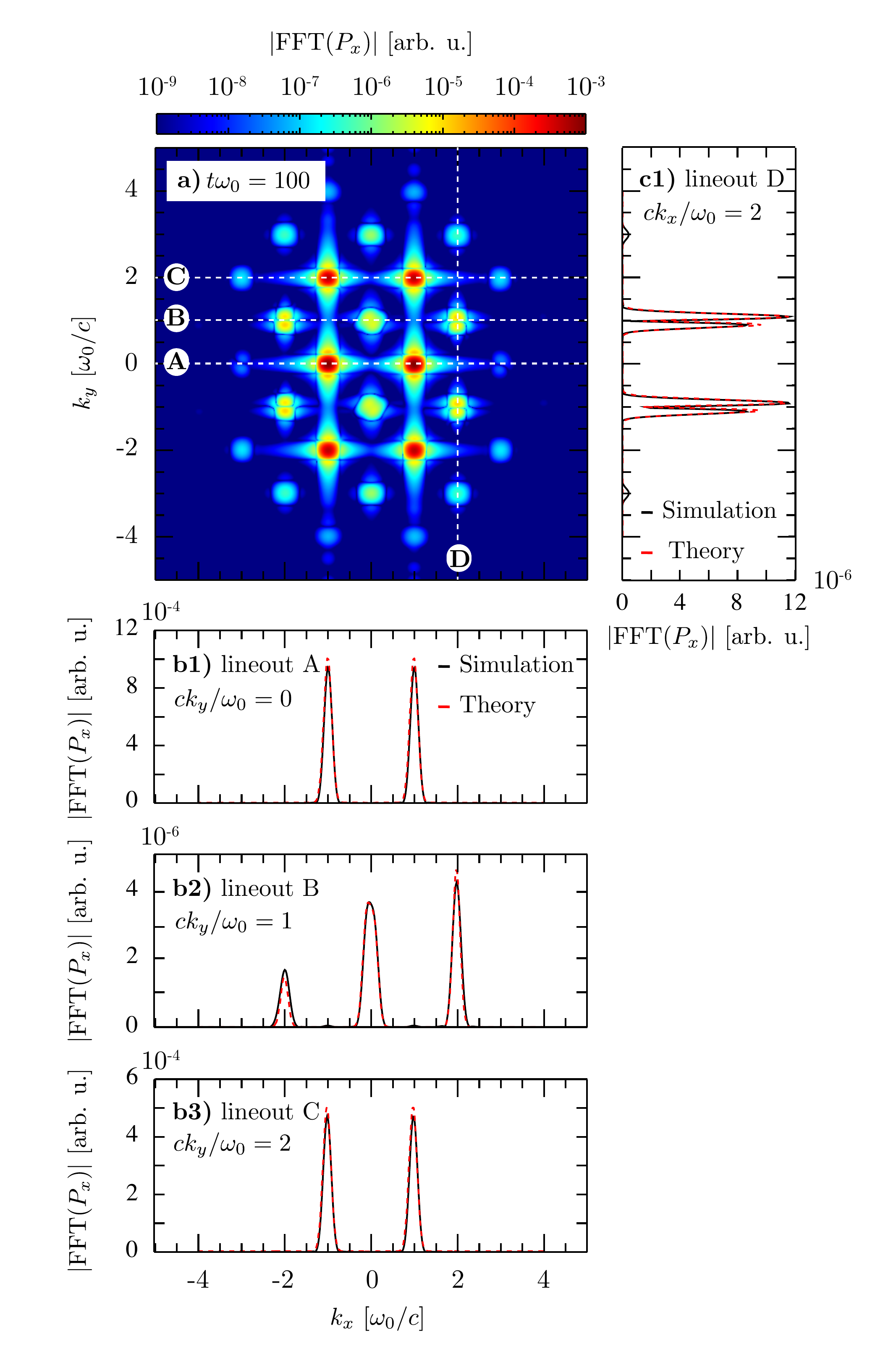}
\caption{Spatial Fourier transform for E$_2$ field at different stages of interaction. (a) Fourier transform of the polarisation $P_x$. The inset b1, b2 , b3 and c1 corresponds to the lineouts A, B, C and D. The red curves represent the Fourier Transform of the theoretical polarisation $P_x$ in equation~\eref{eq:polx}.}
\label{fig:fig_13}
\end{figure}
In our setup, we refer to the index 1 for the pulse propagating in the $x$ direction with wavenumber $k_x$ and to the index 2 for the pulse propagating in the $y$ direction with wavenumber $k_y$. The electromagnetic fields for the pulse 2 can be found by rotating by 90$^\circ$ the fields described for the pulse 1. The non vanishing invariant is 
\begin{eqnarray}
\label{eq:inv_ana}
\mathcal{F} &=& E^2- B^2 \\
\nonumber
&=& \mathcal{F}_x +\mathcal{F}_y - \mathcal{F}_z,
\end{eqnarray}
where $\mathcal{F}_x = (E_{x1}+E_{x2})^2$, $\mathcal{F}_y = (E_{y1}+E_{y2})^2$, and $\mathcal{F}_z = (B_{z1} +B_{z2})^2$. In order to highlight the harmonics generated during the interaction of the two pulses, we use the simplified notation $(n,m) \equiv e^{i(nk_xx+mk_yy)}$ with $n,m \in \mathbb{Z}$. In Fourier space, the invariant is symmetric with respect to the $k_x$ and $k_y$ axes and we can thus just consider for the sake of simplicity one quadrant of the $k$ space. Keeping only the terms for which $k_x,k_y > 0$, the polarization $\boldsymbol{P} = 4\xi\mathcal{F}\boldsymbol{E}$ calculated at first-order reads
\numparts
\begin{eqnarray}
\label{eq:polx}
P_{x} &= \bar{P}_{10}^x(1,0) + \bar{P}_{01}^x(0,1) + \bar{P}_{03}^x(0,3) \nonumber \\
    &+ \bar{P}_{30}^x(3,0)+ \bar{P}_{12}^x(1,2) + \bar{P}_{21}^x(2,1) \\
\label{eq:poly}    
P_{y} &= \bar{P}_{10}^y(1,0) + \bar{P}_{01}^y(0,1) + \bar{P}_{03}^y(0,3) \nonumber \\
    &+ \bar{P}_{30}^y(3,0)+ \bar{P}_{12}^y(1,2) + \bar{P}_{21}^y(2,1).    
\end{eqnarray}
\endnumparts
The full expression for the invariants $\mathcal{F}_{x}, \mathcal{F}_{y}, \mathcal{F}_{z}$ and the polarisation coefficient $\bar{P}_{ij}^{x}, \bar{P}_{ij}^{y}$ can be found in the appendix \ref{appendix}. 
The time at which the interaction is strongest occurs at the full overlap of the two pulses, $\omega_0 t = 100$ for our configuration. The Fourier transform of the polarisation $P_x$ at that time is shown in \fref{fig:fig_13}(a). The harmonics predicted theoretically in equation~\eref{eq:polx} can be readily identified as well as new harmonics such as 
$\bar{P}_{23}^x$ or $\bar{P}_{32}^x$ that result from highest order coupling. The first order harmonics of largest amplitude are $\bar{P}_{10}^x$, $\bar{P}_{12}^x$ (four others are just the symmetric harmonics with respect to the $k_x$ and $k_y$ axis) and are proportional to $\xi(k_0A_0)^3$ whilst $\bar{P}_{01}^x$, $\bar{P}_{21}^x$ scale as $\xi s(k_0A_0)^3$. The harmonics of lowest amplitude $\bar{P}_{30}^x$ do not arise from the interaction of the two pulses but from their self-interaction and are thus proportional to $\xi(sk_0A_0)^3$. A more precise comparison between the simulation and the first order theoretical harmonics of the polarisation $P_x$ is shown on \fref{fig:fig_13}(b1-b2-b3-c1). One notes the very good agreement both for the respective amplitude of the harmonics and their shapes in Fourier space. The theoretical calculation of the first order fields can be carried out by convoluting the 2D Green function of the wave propagator with the linearised source term \cite{matterless_slit, Dipiazza_HHG, DiPiazza} comprised of partial derivatives of the first order polarisation and magnetisation. We have plotted in \fref{fig:fig_14} the temporal evolution of the Fourier transform of the electric field $E_x$. At $t = 0$, the two pulses are fully separated which implies that no interaction has started yet. As a result, only the central wavelength of each pulse is visible, $k_x$ for the longitudinal field of pulse 1 and $k_y$ corresponding to the transverse component of pulse 2. When the two pulses fully overlap, one identifies several harmonics that are identical to the ones we have described for the polarization, which is a direct consequence of the linear property of the wave operator (the polarisation being the source term of the wave equation). Nonetheless, the relative amplitude between the harmonics of the electric field differ from the one we have observed for the polarisation. This is somehow obvious since we are showing here the total electric field $E_x$ which is the sum of all corrections due to the polarisation and magnetisation of the vacuum. Finally, at time $\omega_0 t = 200$, the two pulses have left the zone of interaction, thus ceasing to feed the nonlinear interaction between them. The remainder contributions stem from the self-interaction alone \cite{fedotov}. As also identified by other authors \cite{King1,domenech}, off-axis contributions dominate the spectral region during the overlap and occur due to the field spatio-temporal inhomogeneities. These contributions rapidly fade away with the separation of both pulses. The harmonics that persist after the two pulses have fully separated stem from the self-interaction of each pulse. The amplitude of these harmonics appears to fade away contrary to the 1D case for which the amplitude remains constant. In a 1D simulation, the amplitude of the harmonics generated during the interaction of two pulses does not decrease as they move out of the interaction zone, as seen before whereas in 2D the amplitude of the signal goes down as $1/r$ (and $1/r^2$ in 3D).

The Fourier spectra obtained in these two setups show that the harmonics generated in either case are distinct, thus allowing to clearly distinguish both cases. Future work will include the analytical study of the relative intensity and spectral width of the generated harmonics and their possible relation with other beam parameters. Namely, it is of great interest to understand how the production of these higher harmonics from vacuum may be optimized in terms of the duration of the pulses as these results can provide signatures of experimental relevance. A future setup to explore will also include the interaction of two laser beams at an arbitrary angle $0 < \theta < \frac{\pi}{2}$ to model realistic experimental conditions. If this angular dependence of the interaction is well understood, one could in principle determine how well aligned two ultra-intense beams are by examinating at the Fourier spectrum after a vacuum interaction. Finally, the theoretical predictions, made in the case of two intense focused beams overlapping, on photon merging/splitting \cite{Gies} and four wave mixing \cite{King-Keitel} could also be verified with an extension of this present code in 3D dimensions, which is computational very demanding but still feasible. A straightforward extension to the code could also be the addition of higher order terms in the Euler Heisenberg Lagrangian. This would allow us to explore the shock formation and asymptotic field generated via six-, eight- or higher wave mixing processes \cite{domenech,King1}.

\begin{figure}
\centering
\includegraphics[width=0.75\linewidth]{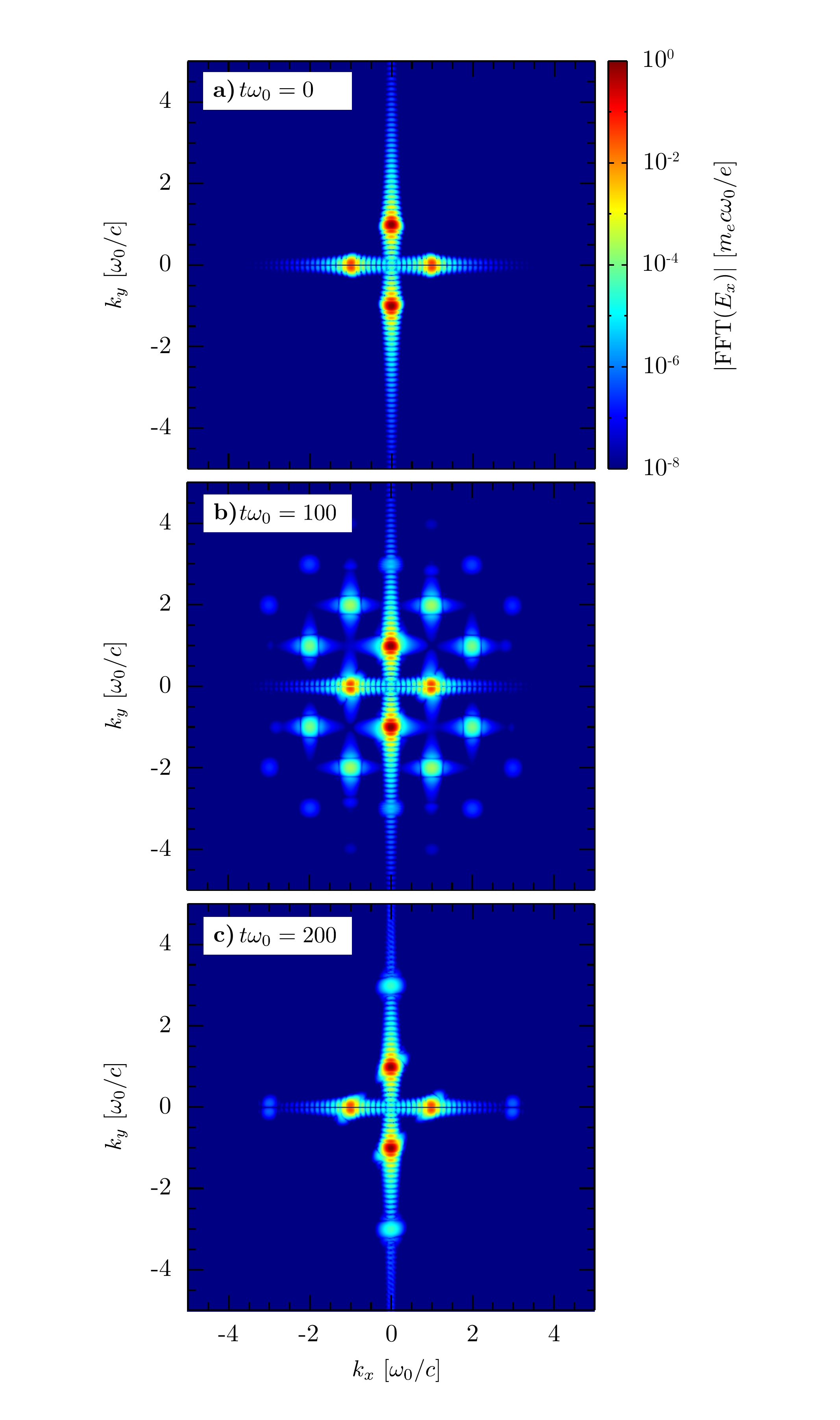}
\caption{Spatial Fourier transform for E$_2$ field at different stages of interaction. (a) Initial Fourier space, if there were no vacuum non-linearities this spectrum would remain unchanged throughout the interaction. (b) At peak of nonlinear interaction when the pulses are completely overlapped in space. (c) Asymptotic state: after the nonlinear interaction the pulses propagate independently but with higher harmonics generated from the interaction.}
\label{fig:fig_14}
\end{figure}

\section{Conclusions}
\label{sec:conclusions}

A numerically stable and robust generalized Yee scheme to solve the nonlinear set of QED Maxwell's equations was developed and incorporated in a standard PIC loop. This work represents an important step  towards modeling plasma dynamics in extreme scenarios when QED processes significantly alter the collective behavior of the system. Furthermore, the algorithm is fully generalizable to include higher order corrections (such as six-photon scattering or higher order terms). These terms are to be included in the future as they have been shown to be necessary to fully simulate certain scenarios \cite{King1}. Our numerical model can be used to design planned experiments, leveraging on ultra-intense laser facilities able to deliver intensities of $10^{23} - 10^{24}$ W/cm$^{2}$, to verify for the first time the dynamics of the quantum vacuum below the Schwinger limit. The simulations confirm predicted optical phenomena such as vacuum birefringence and high harmonics generation in one-dimensional setups with an excellent accuracy. The algorithm was also extended for two-dimensional scenarios where two setups of interacting Gaussian beams were studied. The results highlight the importance of transverse beam effects and hint that the generation of higher harmonics from quantum vacuum can  be achieved via this interaction. The spectrum of the harmonics could provide a direct measurement of important beam properties such as the peak intensity and alignment. This algorithm may also be used to test two and three dimensional setups that have been proposed in the literature (where transverse and finite spot size effects are taken into account under certain approximations), thus complementing the results of previous theoretical works \cite{Dinu1, Dinu2, DiPiazza}.  Finally  our algorithm contributes to the generalization of the Yee scheme, one of the most successful and commonly used  algorithms in computational physics, to scenarios where nonlinear polarization and magnetization can impact EM propagation.

This work is partially supported by the European Research Council (ERC-2015-AdG Grant 695088). RT and FC are supported by FCT (Portugal) (grants PD/BD/142971/2018 and PD/BD/114307/2016, respectively) in the framework of the Advanced Program in Plasma Science and Engineering (APPLAuSE, FCT grant PD/00505/2018). We acknowledge PRACE for granting access to MareNostrum (Barcelona Supercomputing Center, Spain)  where the simulations presented in this work were performed. The author acknowledge useful discussions with A. Di Piazza, M. Marklund, H. Ruhl, and S. Meuren.

\appendix

\section{Harmonics Coefficients}
\label{appendix}
Plugging the expression of the electromagnetic fields of equation~\eref{eq:beam_davis1} into equation~\eref{eq:inv_ana}, we obtain
\numparts
\begin{eqnarray}
\mathcal{F}_x &= \bar{E}_{x1}^2(-2,0)+ 2\bar{E}_{x1}\bar{E}_{x1}^*(0,0) + \bar{E}_{x1}^{*2}(2,0) \nonumber \\
               &+ \bar{E}_{x2}^2(0,-2)+ 2\bar{E}_{x2}\bar{E}_{x2}^*(0,0) + \bar{E}_{x2}^{*2}(0,2) \nonumber \\
               &+ 2\bar{E}_{x1}\bar{E}_{x2}(-1,-1)+ 2\bar{E}_{x1}\bar{E}_{x2}^*(-1,1) + 2\bar{E}_{x1}^*\bar{E}_{x2}^*(1,1) \nonumber \\
               &+ 2\bar{E}_{x1}^*\bar{E}_{x2}(1,-1) \nonumber \\
\mathcal{F}_y &= \bar{E}_{y1}^2(-2,0)+ 2\bar{E}_{y1}\bar{E}_{y1}^*(0,0) + \bar{E}_{y1}^{*2}(2,0) \nonumber \\
               &+ \bar{E}_{y2}^2(0,-2)+ 2\bar{E}_{y2}\bar{E}_{y2}^*(0,0) + \bar{E}_{y2}^{*2}(0,2) \nonumber \\
               &+ 2\bar{E}_{y1}\bar{E}_{y2}(-1,-1)+ 2\bar{E}_{y1}\bar{E}_{y2}^*(-1,1) + 2\bar{E}_{y1}^*\bar{E}_{y2}^*(1,1) \nonumber \\
               &+ 2\bar{E}_{y1}^*\bar{E}_{y2}(1,-1) \nonumber \\ 
\mathcal{F}_z &= \bar{B}_{z1}^2(-2,0)+ 2\bar{B}_{z1}\bar{B}_{z1}^*(0,0) + \bar{B}_{z1}^{*2}(2,0) \nonumber \\
               &+ \bar{B}_{z2}^2(0,-2)+ 2\bar{B}_{z2}\bar{B}_{z2}^*(0,0) + \bar{B}_{z2}^{*2}(0,2) \nonumber \\
               &+ 2\bar{B}_{z1}\bar{B}_{z2}(-1,-1)+ 2\bar{B}_{z1}\bar{B}_{z2}^*(-1,1) + 2\bar{B}_{z1}^*\bar{B}_{z2}^*(1,1) \nonumber \\
               &+ 2\bar{B}_{z1}^*\bar{B}_{z2}(1,-1) \nonumber    
\end{eqnarray}               
\endnumparts

The first quadrant coefficients present in equations~\eref{eq:polx} and \eref{eq:poly} are
\numparts
\begin{eqnarray}
\bar{P}_{10}^x &=  4\xi \Big[\bar{E}_{x1}\left(3\bar{E}^{*2}_{x1}+\bar{E}^{*2}_{y1}-\bar{B}^{*2}_{z1}\right)\nonumber\\ 
               &+ \bar{E}^{*}_{x1}\left(6\bar{E}_{x2}\bar{E}^{*}_{x2}+2\bar{E}_{y1}\bar{E}^{*}_{y1}+2\bar{E}_{y2}\bar{E}^{*}_{y2}-2\bar{B}_{z1}\bar{B}^{*}_{z1}-2\bar{B}_{z2}\bar{B}^{*}_{z2}\right)\nonumber\\
               &+ \bar{E}_{x2}\left(2\bar{E}^{*}_{y1}\bar{E}^{*}_{y2}-2\bar{B}^{*}_{z1}\bar{B}^{*}_{z2}\right)\nonumber\\
               &+ \bar{E}^{*}_{x2}\left(2\bar{E}^{*}_{y1}\bar{E}_{y2}-2\bar{B}^{*}_{z1}\bar{B}_{z2}\right)\Big] \nonumber\\
\bar{P}_{01}^x &=  4\xi \Big[\bar{E}_{x1}\left(6\bar{E}^{*}_{x1}\bar{E}^{*}_{x2}+2\bar{E}^{*}_{y1}\bar{E}^{*}_{y2}-\bar{B}^{*}_{z1}\bar{B}^{*}_{z2}\right)\nonumber\\ 
               &+ \bar{E}^{*}_{x1}\left(2\bar{E}_{y1}\bar{E}^{*}_{y2}-2\bar{B}_{z1}\bar{B}^{*}_{z2}\right)\nonumber\\
               &+ \bar{E}_{x2}\left(3\bar{E}^{*2}_{x2}+\bar{E}^{*2}_{y2}-\bar{B}^{*2}_{z2}\right)\nonumber\\
               &+ \bar{E}^{*}_{x2}\left(2\bar{E}_{y1}\bar{E}^{*}_{y1}+2\bar{E}_{y2}\bar{E}^{*}_{y2}-2\bar{B}_{z1}\bar{B}^{*}_{z1}-2\bar{B}_{z2}\bar{B}^{*}_{z2}\right)\Big] \nonumber\\
\bar{P}_{03}^x &= 4\xi\bar{E}^{*}_{x2}\Big[\bar{E}_{x2}^{*2} + \bar{E}_{y2}^{*2} - \bar{B}_{z2}^{*2}\Big] \nonumber \\  
\bar{P}_{30}^x &= 4\xi\bar{E}^{*}_{x1}\Big[\bar{E}_{x1}^{*2} + \bar{E}_{y1}^{*2} - \bar{B}_{z1}^{*2}\Big] \nonumber \\    
\bar{P}_{12}^x &= 4\xi\Big[\bar{E}^{*}_{x1}\left(3\bar{E}_{x2}^{*2} + \bar{E}_{y2}^{*2} - \bar{B}_{z2}^{*2}\right) \nonumber \\  
             &+ \bar{E}^{*}_{x2}\left(2\bar{E}_{y1}^{*}\bar{E}_{y2}^{*}-2\bar{B}_{z1}^{*}\bar{B}_{z2}^{*}\right)\Big] \nonumber \\ 
\bar{P}_{21}^x &= 4\xi\Big[\bar{E}^{*}_{x1}\left(3\bar{E}_{x1}^{*}\bar{E}_{x2}^{*} + 2\bar{E}_{y1}^{*}\bar{E}_{y2}^{*} - 2\bar{B}_{z1}^{*}\bar{B}_{z2}^{*}\right) \nonumber \\  
             &+ \bar{E}^{*}_{x2}\left(\bar{E}_{y1}^{*2}-\bar{B}_{z1}^{*2}\right)\Big] \nonumber                                                 
\end{eqnarray}
\endnumparts

\numparts
\begin{eqnarray}
\bar{P}_{10}^y &=  4\xi \Big[\bar{E}_{y1}\left(\bar{E}^{*2}_{x1}+3\bar{E}^{*2}_{y1}-\bar{B}^{*2}_{z1}\right) \nonumber \\
                &+ \bar{E}^{*}_{y1} \left(2\bar{E}_{x1}\bar{E}^{*}_{x1}+2\bar{E}_{x2}\bar{E}^{*}_{x2}+6\bar{E}_{y2}\bar{E}^{*}_{y2}-2\bar{B}_{z1}\bar{B}^{*}_{z1}-2\bar{B}_{z2}\bar{B}^{*}_{z2}\right)\nonumber \\
                &+ \bar{E}_{y2} \left(2\bar{E}^{*}_{x1}\bar{E}^{*}_{x2}-2\bar{B}^{*}_{z1}\bar{B}^{*}_{z2}\right)\nonumber \\
                &+ \bar{E}^{*}_{y2} \left(2\bar{E}^{*}_{x1}\bar{E}_{x2}-2\bar{B}^{*}_{z1}\bar{B}_{z2}\right)\Big]\nonumber \\
\bar{P}_{01}^y &=  4\xi \Big[\bar{E}_{y1}\left(2\bar{E}^{*}_{x1}\bar{E}^{*}_{x2}+6\bar{E}^{*}_{y1}\bar{E}^{*}_{y2}-2\bar{B}^{*}_{z1}\bar{B}^{*}_{z2}\right) \nonumber \\
                &+ \bar{E}^{*}_{y1} \left(2\bar{E}_{x1}\bar{E}^{*}_{x2}-2\bar{B}_{z1}\bar{B}^{*}_{z2}\right)\nonumber \\
                &+ \bar{E}_{y2} \left(\bar{E}^{*2}_{x2}+3\bar{E}^{*2}_{y2}-\bar{B}^{*2}_{z2}\right)\nonumber \\
                &+ \bar{E}^{*}_{y2} \left(2\bar{E}_{x1}\bar{E}^{*}_{x1}+2\bar{E}_{x2}\bar{E}^{*}_{x2}-2\bar{B}_{z1}\bar{B}^{*}_{z1}-2\bar{B}_{z2}\bar{B}^{*}_{z2}\right)\Big]\nonumber \\
\bar{P}_{03}^y &= 4\xi\bar{E}_{y2}^{*}\Big[\bar{E}_{x2}^{*2} + \bar{E}_{y2}^{*2} - \bar{B}_{z2}^{*2}\Big] \nonumber \\               
\bar{P}_{30}^y &= 4\xi\bar{E}_{y1}^{*}\Big[\bar{E}_{x1}^{*2} + \bar{E}_{y1}^{*2} - \bar{B}_{z1}^{*2}\Big] \nonumber \\
\bar{P}_{12}^y &= 4\xi\Big[\bar{E}_{y1}^{*} \left(\bar{E}_{x2}^{*2}+3\bar{E}_{y2}^{*2}-\bar{B}_{z2}^{*2}\right)\nonumber \\
                &+ \bar{E}_{y2}^{*} \left(2\bar{E}_{x1}^{*}\bar{E}_{x2}^{*}-2\bar{B}_{z1}^{*}\bar{B}_{z2}^{*}\right)\Big]\nonumber \\
\bar{P}_{21}^y &=  4\xi\Big[\bar{E}_{y1}^{*}\left(2\bar{E}_{x1}^{*}\bar{E}_{x2}^{*}+3\bar{E}_{y1}^{*}\bar{E}_{y2}^{*}-2\bar{B}_{z1}^{*}\bar{B}_{z2}^{*}\right)\nonumber \\
                &+ \bar{E}_{y2}^{*} \left(\bar{E}_{x1}^{*2}-\bar{B}_{z1}^{*2}\right)\Big]\nonumber
\end{eqnarray}
\endnumparts

\section*{References}
\bibliography{references}

\end{document}